\documentclass[a4paper,11pt]{article}
\usepackage{jheppub} \pdfoutput=1
\usepackage[utf8]{inputenc} %Support for additional characters in bibliography
\usepackage{amsfonts,amssymb, mathrsfs, amsmath, esint,bm,enumitem}
\allowdisplaybreaks
\usepackage{siunitx}
\usepackage{bookmark}
\usepackage{subcaption}
\usepackage{lipsum}
\usepackage{ragged2e}
\DeclareCaptionJustification{justified}{\justifying}
\usepackage{diagbox}
\usepackage{latexsym}
\usepackage{graphicx}
\usepackage[dvipsnames,table]{xcolor}
\usepackage{booktabs,multirow}
\usepackage[capitalize]{cleveref}%nameinlink to have clickable "Eq."
\usepackage{physics,braket}
\usepackage{comment}
\usepackage[normalem]{ulem}
\graphicspath{{./Figures/}}%{../2_Mathematica/}{../2_Mathematica/1_warm_2}{../2_Mathematica/1_cold_2}}

\usepackage{tikz}
\usetikzlibrary{decorations.pathmorphing}
\usetikzlibrary{math}
\usetikzlibrary{positioning,calc}

\usepackage{url}
\renewcommand{\d}{\mathrm{d}}
\newcommand{\parfrac}[2]{\left(\frac{#1}{#2}\right)}

\newcommand{\MeV}{\,\text{MeV}}
\newcommand{\GeV}{\,\text{GeV}}

\newcommand{\thi}{\theta_\text{i}}
\newcommand{\dth}{\delta\theta}
\newcommand{\sth}{\sigma_\theta}
\newcommand{\dthpr}{\dth_\text{(par.res.)}}
\newcommand{\fa}{f_a}
\newcommand{\fainf}{\fa^\text{(inf)}}
\newcommand{\HI}{H_I}
\newcommand{\He}{H_e}
\renewcommand{\ae}{a_e}
\newcommand{\te}{t_e}
\newcommand{\ke}{k_e}
\newcommand{\kend}{\ke}
\newcommand{\TRHmax}{T_\mathrm{inst.\,RH}}
\newcommand{\Tmax}{T_\text{max}}
\newcommand{\gs}{g_\star}
\newcommand{\biso}{\beta_\text{iso}}
\newcommand{\Pa}{\mathcal P_a}
\newcommand{\Sag}{\mathcal S_{a\gamma}}
\newcommand{\As}{A_s}
\newcommand{\vx}{\mathbf{x}}
\newcommand{\vk}{\mathbf{k}}
\newcommand{\fanh}{\mathscr{F}_\text{anh.}}
\newcommand{\NCMB}{N_\textsc{cmb}}
\renewcommand{\l}{\lambda_\Phi}
\newcommand{\mphi}{m_\Phi}
\newcommand{\as}{\alpha_\text{s}}

\renewcommand{\epsilon}{\varepsilon}
\newcommand{\MP}{M_\textsc{p}}
\newcommand{\UPQ}{U(1)_\textsc{pq}}
\newcommand{\ovX}{\overline X}
\newcommand{\ovY}{\overline Y}
\newcommand{\Gs}{\Gamma_\sigma}
\newcommand{\ms}{m_\sigma}
\newcommand{\yKSVZ}{y_\textsc{ksvz}}
\newcommand{\Gsph}{\Gamma_\text{sph}}
\newcommand{\Usph}{\Upsilon_\text{sph}}
\newcommand{\TDR}{T_\textsc{dr}}
\newcommand{\rDR}{\rho_\textsc{dr}}

\definecolor{constant}{rgb}{0.5, 0.5, 0.65}
\subheader{ZU-TH 38/25}

\title{Revisiting Isocurvature Bounds on the Minimal QCD Axion}

\author[a,b]{Peter W.~Graham,}
\author[a,c,d]{Davide~Racco}
\affiliation[a]{Stanford Institute for Theoretical Physics, Stanford University, 382 Via Pueblo Mall, Stanford, CA 94305, U.S.A.}
\affiliation[b]{Kavli Institute for Particle Astrophysics \& Cosmology, Department of Physics, Stanford University, Stanford, California 94305, USA}
\affiliation[c]{Physik-Institut, Universit\"at Z\"urich, Winterthurerstrasse 190, 8057 Z\"urich, Switzerland}
\affiliation[d]{Institut f\"ur Theoretische Physik, ETH Z\"urich,Wolfgang-Pauli-Str.\ 27, 8093 Z\"urich, Switzerland}
\emailAdd{pwgraham@stanford.edu}
\emailAdd{davide.racco@uzh.ch}
\abstract{
The QCD axion has important connections to early universe cosmology.  For example, it is often said that isocurvature limits rule out a combination of high axion decay constant, $f_a$, and high inflationary Hubble scale, $H_I$.
High scales are theoretically motivated, so it is important to ask how robust this constraint is.
We demonstrate that this constraint is naturally evaded when the quartic coupling of the complex $U(1)_\textsc{pq}$-breaking field is small (e.g.~$\lesssim 10^{-6}$).
In this case, $f_a$ changes from a larger value during inflation to a smaller value in the later universe, suppressing isocurvature perturbations.
Importantly, we show that in large parts of parameter space this solution is not jeopardised by overproduction of the axion through parametric resonance.
The isocurvature bounds are thus dependent on UV physics.  We have found that, even for the minimal QCD axion, large parts of UV parameter space at both high $f_a$ and high $H_I$ are in fact allowed, not ruled out by isocurvature constraints.
}

\begin{document}

\maketitle

% !TEX root = Draft_Axion-Isocurvature.tex

\section{Introduction}
\label{sec:intro}

\subsection{Background}

The QCD axion is one of the best theoretically-motivated extensions to the Standard Model of particle physics (see e.g.~\cite{Graham:2015ouw,Baryakhtar:2025jwh} for reviews).  It solves the strong CP problem and also gives a natural dark matter candidate.  The question of the production of axion dark matter then becomes a critical one.  Interestingly, the axion dark matter abundance today is sensitive to initial conditions for the axion field arising from inflation at the beginning of the universe.  This means that cosmological measurements of inflation may affect possibilities for axion dark matter and similarly any direct observations of axion dark matter could inform our understanding of the beginning of the universe.

There are two main scenarios for axion dark matter production in the early universe, the pre-inflationary and post-inflationary regimes.  In the post-inflationary regime, the PQ symmetry breaks sometime after inflation, giving rise to the axion degree of freedom.  In this case the axion field begins initially spread over its entire range (an effective $\theta$-angle from 0 to $2\pi$).  Since the initial condition is fixed, this produces a sharp prediction for the axion dark matter abundance as a function of the axion mass, and thus only a single axion mass will produce the observed dark matter abundance.  This scenario can produce axion domain walls and cosmic strings.

In the pre-inflationary regime, PQ symmetry is already broken before the end of inflation.  In this case, we generically expect that the axion field will initially have a (roughly) homogeneous value, $\thi$, over the entire observable universe.  Then in order to produce the observed dark matter abundance for a given $\thi$, a particular value for the axion decay constant $f_a$ (or equivalently the axion mass $m_a$) must be chosen which satisfies \cref{eq:theta_i}.  This means that at least one of the two dimensionless constants of the theory $\thi$ and $\frac{f_a}{\MP}$ must be small.  
UV arguments motivate considering a wide range for $f_a$ (see e.g.~\cite{Svrcek:2006yi,Demirtas:2020dbm, Gendler:2023kjt,Gendler:2024adn, Petrossian-Byrne:2025mto, Loladze:2025uvf, Benabou:2025kgx}), including large values which would then require a small $\thi$.  This small $\thi$ can occur naturally for the QCD axion with a long period of inflation (the `stochastic axion' scenario) \cite{Graham:2018jyp, Takahashi:2018tdu}, and additionally is explained naturally in many models (see e.g.~\cite{Agrawal:2017eqm, Nomura:2015xil, Dine:1982ah, Steinhardt:1983ia, Lazarides:1990xp, Kawasaki:1995vt, Dvali:1995ce, Choi:1996fs, Banks:1996ea, Banks:2002sd, Bao:2022hsg}).

There are restrictions on the allowed parameter space for the pre-inflationary regime due to isocurvature constraints (see e.g.~\cite{Lyth:1991ub, Fox:2004kb, Hertzberg:2008wr, Visinelli:2009zm, Chen:2023txq}).  The value of the axion field will acquire quantum fluctuations during inflation.  In the later universe these become fluctuations of the axion density which are uncorrelated with the regular adiabatic density perturbations, and hence this gives rise to isocurvature fluctuations.  The region of parameter space which was claimed to be excluded by isocurvature constraints is shown in \cref{fig:baseline isocurvature}.
Note that it is possible to build models to evade these isocurvature constraints (see e.g.~\cite{Jeong:2013xta, Nomura:2015xil, Kawasaki:2015lpf, Takahashi:2015waa, Harigaya:2015hha, Nakayama:2015pba, Ballesteros:2021bee, Choi:2014uaa,Kawasaki:2014una}).
However the minimal model for the QCD axion plus inflation, which is what we consider here, is claimed to be constrained by isocurvature.
As is apparent in the figure, larger Hubble scales during inflation are generally claimed to be excluded by these constraints for the pre-inflationary axion scenario.

\subsection{Executive Summary}

\begin{figure}[t!]\centering
\includegraphics[width=.9\textwidth, trim = 160 460 2 2, clip]{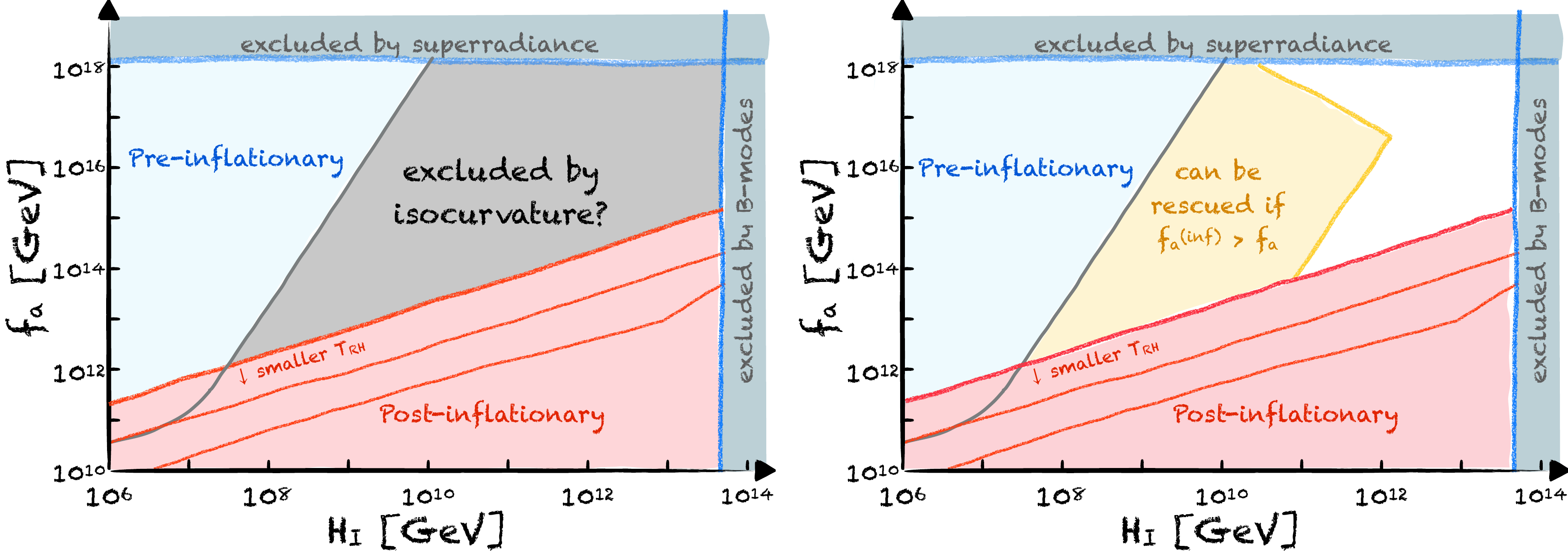}
\caption{A rough sketch of axion parameter space in the axion decay constant $\fa$ vs the inflationary Hubble scale $\HI$.  
The left figure shows the region of parameter space which has conventionally been claimed to be ruled out by isocurvature constraints (grey).  
However we demonstrate in this paper that a significant fraction of that region is in fact allowed, shown as the yellow region in the right plot.  Whether the isocurvature fluctuations are too large in this region is determined by parameters of the underlying PQ-breaking potential and the model of inflation and reheating.
Both figures show the parts of parameter space that would be in the pre-inflationary (blue) and post-inflationary (red) regimes, as well as bounds from B-modes and superradiance (light grey).}
\label{fig:sketch}
\end{figure}

We demonstrate in this paper that in fact a large portion of the supposedly excluded region is actually allowed, as shown schematically in \cref{fig:sketch}.  We do not add anything to the model in order to avoid the isocurvature constraints.  We show that this region is allowed for a simple QCD axion with no other fields added to the model, in a certain part of axion parameter space.  In particular, we show that if there is one small parameter in the UV theory of the complex scalar from which the axion arises (whose absolute value we call the saxion field) then the axion in fact does not acquire isocurvature fluctuations in the region shown in \cref{fig:sketch}.

Isocurvature fluctuations can be suppressed in this regime due to a dynamical change to the saxion potential during inflation.  The saxion may have a coupling to the Ricci scalar which causes an increase in its VEV during inflation.  This means the effective axion decay constant during inflation, $\fainf$, is much larger than the `true' $f_a$ later in the universe.  The Hubble-sized quantum fluctuations of the axion field during inflation are divided by this larger $\fainf$, thus producing smaller overall fluctuations in $\thi$ and so suppressing isocurvature.
As a concrete example in this paper we consider the saxion coupling to the Ricci scalar since it is naturally allowed and should be present.  However other UV couplings, e.g.~couplings of the saxion to the inflaton, may also produce a similar effect.

In order for this mechanism to work, there must be one small parameter in the UV Lagrangian of the saxion.  The ratio of the saxion quartic $\l$ to its gravitational coupling $\xi$ must be small (though note that $\xi$ can be $\mathcal{O}(1)$).
The exact values of these parameters are dependent on the UV models of inflation and the saxion.  For the particular example we consider in most detail, the quartic must be $\l \lesssim 10^{-6}$ in order to start cutting into the isocurvature bound.
The smaller this ratio $\l / \xi$ is, the larger a region can be rescued from isocurvature constraints (for reasons discussed below we choose to cut off our scan at a lower limit of $\l = 10^{-18}$ though one could have continued lower).
There are known models which naturally have a small effective quartic or very flat potential for the saxion, e.g.~in supergravity~\cite{Moxhay:1984am,Abe:2001cg,Nakamura:2008ey,DEramo:2015iqd}. The kinetic (or acoustic) misalignment mechanism tends to rely on this as well in order to enhance the energy density associated to the axion rotation (see e.g.~\cite{Co:2017mop, Co:2019jts, Co:2020dya, Domcke:2022wpb, Eroncel:2022vjg, Eroncel:2024rpe,Bodas:2025eca, Eroncel:2025qlk}).  Kinetic misalignment is also similar in often using a negative, Hubble-induced mass during inflation which drives the saxion to large field values (giving a large $\fa$ during inflation).
Other similar mechanisms have been proposed with a flat potential and a large initial saxion field value, for example to generate the baryon asymmetry from the axion \cite{Co:2019wyp, Co:2023mhe}.

The possibility of reducing isocurvature constraints through such a mechanism has previously been considered \cite{Linde:1991km,Kasuya:1996ns,Kawasaki:2008jc,Folkerts:2013tua, Fairbairn:2014zta, Kearney:2016vqw, Kawasaki:2013iha,% lattice study assuming MD after infl
Harigaya:2015hha,%parametric resonance with attractor dynamics of saxion
Kobayashi:2016qld,%with N_DW=1, the strings overclose
Co:2017mop,Co:2020dya,%axion DM after parametric resonance
Ballesteros:2021bee}.  However it was worried that parametric resonance may generate too large an axion abundance, ruling out this scenario.  We demonstrate that in fact there is a significant amount of viable parameter space where isocurvature bounds are removed and yet parametric resonance is not strong enough to overproduce the axion.  As we show, the strength of the parametric resonance depends on many detailed aspects of the UV scenario, including the model of inflation and reheating and the saxion potential.  Such UV parameters are hidden on the simple plots of $\HI$ vs $\fa$ as in \cref{fig:sketch}, but in fact determine whether the axion model will be ruled out by isocurvature.
For example, in different models of inflation and reheating the transition in PQ breaking scale from the high scale $\fainf$ down to the scale in the late universe $\fa$ can occur at different rates.  The more adiabatic this transition is, the weaker the parametric resonance is.
We investigate a couple different inflation models and display the allowed parts of parameter space in Fig.~\ref{fig:isocurvature-rescued}.  Considering other inflation/reheating models would likely expand the allowed region even further.

Ultimately, whether a point in the $\HI$ vs $\fa$ plane is ruled out by isocurvature or not is a UV-dependent question.  There are UV realizations of the minimal QCD axion and inflation in which significant parts of the parameter space of the claimed isocurvature bound are in fact allowed and do not overproduce isocurvature perturbations.

% !TEX root = Draft_Axion-Isocurvature.tex

\section{Standard isocurvature constraints on axion dark matter}
\label{sec:axion-iso}

In this section, we concisely review the standard constraints on axion dark matter (DM) from isocurvature perturbations.
According to CMB measurements, matter curvature perturbations spatially track the perturbations of SM radiation $\gamma$ around the time of matter-radiation equality. 
Any deviation for a species $a$ is parametrised by \textit{isocurvature} perturbations $\Sag(\vx) = 3(\zeta_a(\vx)-\zeta_\gamma(\vx))$, where $\zeta_i(\vx)=\Psi(\vx)-H \delta\rho_i(\vx)/\dot \rho_i$ is the curvature perturbation of the species $i=a,\gamma$ \cite{Kodama:1984ziu}.
The strongest observational bound%
\footnote{Other interesting signatures of dark matter isocurvature may appear in correlators of large-scale structures \cite{Chen:2023txq}.} 
on isocurvature perturbations on large scales comes from the CMB measurements of Planck18 \cite[Table~14]{Planck:2018jri}.%
\footnote{The very recent results by ACT \cite[Table~6]{ACT:2025tim} are mildly weaker, with Planck18 + ACT giving $\biso < 0.042$, and Planck18 + ACT + DESI 1y BAO yielding $\biso < 0.049$.}
The limit on the power spectrum $\Pa$ of isocurvature perturbations $\Sag(\vx)$ that is relevant to our discussion is
\begin{equation}
\Pa < \As \biso\,, \quad \As = 2.1\cdot 10^{-9}, \quad\quad \biso = 0.038\,,
\end{equation}
where $\As$ is the amplitude of scalar perturbations measured at the CMB pivot scale, and $\biso$ corresponds to the case of uncorrelated isocurvature perturbations (``axion I'').

We now specialise the discussion to isocurvature perturbations of axion DM, which we can compute at a time during the early period of radiation domination when the axion has started oscillating around its minimum, so that its equation of state is that of non-relativistic matter, $\rho_a(\vx) = m\, n_a(\vx)$ and $\dot \rho_a = -3 H\rho_a$. 
We consider the spatially flat gauge $\Psi=0$. 
At early times, deep in radiation domination, the axion is a tiny component of the total energy density, and $\delta \rho_a/\rho_a \gg \delta \rho_a/\rho_\gamma \sim \delta \rho_\gamma/\rho_\gamma$, so that we can approximate $\Sag= 3\zeta_a$. 
We can then write the contribution from the axion field to $\Pa$ as \cite{Hertzberg:2008wr,Visinelli:2009zm} %p.6 
\begin{equation}
\label{eq:delta n_a}
\Pa = \left\langle \parfrac{\delta n_a}{n_a}^2 \right\rangle,\quad n_a(\vx)=\frac{\rho_a(\vx)}{m_a} = \frac 12 m_a \fa^2 \theta(\vx)^2\, \fanh\big(\theta(\vx)\big)
\end{equation}
where $\theta(x) \equiv a(x)/\fa$, while $\fanh(\theta)$ accounts for anharmonicity effects and differs from 1 only for $\theta \gtrsim 2$. In this work, we are interested in a region of parameter space where $\theta<1$, so we ignore anharmonic effects for simplicity (although we show the correct isocurvature curve in our plots including those effects, see e.g.~\cite{Kobayashi:2013nva}). 
We define the average value $\thi\equiv \langle \theta(x)\rangle$ for the initial misalignment angle in our observable Universe, and the spatial fluctuations $\dth(\vx) \equiv \theta(\vx)-\thi$. 
We assume that $\dth$ is a Gaussian variable, with standard deviation set by quantum fluctuations during inflation
\begin{equation}
\label{eq:sigma theta}
\sth \equiv \sqrt{\langle \dth(\vx)^2\rangle} 
  = \frac{\HI}{2\pi \fainf}\sqrt{\NCMB} \,.
\end{equation}
Here, $\fainf$ is the vev of the radial direction of the complex PQ field (which we dub the \textit{saxion}) during inflation, that we allow to differ from its present-day value $\fa$.
The parameter $\NCMB\gtrsim 1$ quantifies the growth of the variance of the axion field during the range of $e$-folds contributing to the experimental constraints on isocurvature, and we take $\NCMB=1$ for simplicity.%
\footnote{Previous analyses provided arguments for e.g.~$\sqrt{\NCMB}=2$ \cite{Hertzberg:2008wr} or 4 \cite{Lyth:1991ub}.}

We can now write the fluctuations in the axion number density  (neglecting anharmonic effects and treating the axion potential as quadratic, which is appropriate for $\theta<1$) in terms of the initial fluctuations when the axion field started oscillating,
\begin{equation}
\frac{\delta n_a}{n_a} \approx \frac{\theta^2 -\langle \theta^2\rangle}{\langle \theta^2\rangle} 
  = \frac{\big(\thi^2 +2\thi \delta\theta + (\delta\theta)^2 \big) 
  - \big(\thi^2+\sth^2 \big) }{\thi^2+\sth^2}
  = \frac{2\thi \delta\theta + (\delta\theta)^2 - \sth^2}{\thi^2+\sth^2} \,.
\end{equation}
The relevant figure for isocurvature perturbations is the variance of these fluctuations,
\begin{equation}
\left\langle\parfrac{\delta n_a}{n_a}^2\right\rangle 
  \approx \frac{\langle \dth^4 +4\thi^2 (\dth)^2 +\sth^4 -2(\dth)^2\sth^2 +(\text{odd in }\dth) \rangle}{(\thi^2 +\sth^2)^2}
  = \frac{4\thi^2\sth^2+2\sth^4}{(\thi^2 +\sth^2)^2} \overset{\thi\gg\sth}{\longrightarrow}
  \frac{4\sth^2}{\thi^2}
\,, 
\end{equation}
where we assumed that the axion fluctuations are Gaussian.
We can then rewrite the isocurvature constraint as (in the limit $\thi \lesssim 2$, and using that $\sth\ll\thi$ with this constraint)
\begin{equation}
\label{eq:isocurvature bound}
\frac{\HI}{\fainf\thi} < \sqrt{\frac{\pi^2\As \biso}{\NCMB}} = 2.8\cdot 10^{-5}\,.
\end{equation}
The initial misalignment $\thi$ yielding the DM abundance for the axion field is in one-to-one correspondence with the axion decay constant $\fa$, and encodes various features of the axion physics at the epoch of the QCD phase transition.
In the dilute instanton gas approximation for the axion potential (approximated to a quadratic potential for $\thi\lesssim 2$), with the simplification that the axion starts oscillating exactly at $m_a = 3H(T_\text{osc})$, assuming that the temperature-dependent axion mass scales as $m_a(T)=0.026\, m_a (160 \MeV/T)^4$, one finds \cite{GrillidiCortona:2015jxo,DiLuzio:2020wdo}
\begin{equation}
\label{eq:theta_i}
\thi = \parfrac{7.4\cdot 10^{11}\GeV}{\fa}^{7/12}
\end{equation}%
\begin{figure}[t]\centering
\includegraphics[width=.8\textwidth]{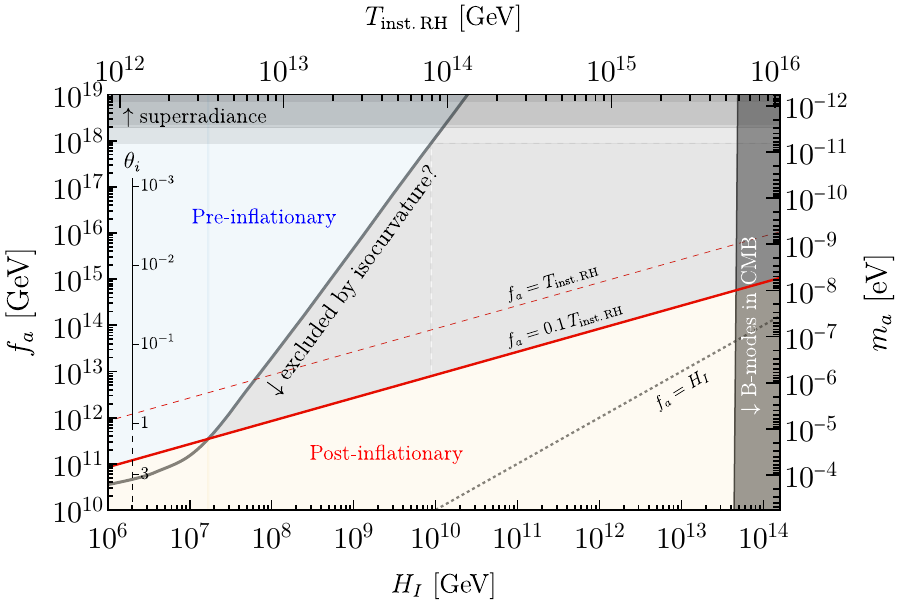}\caption{
Parameter space in the plane $(\HI,\fa)$ where the QCD axion can be the DM respecting the \textcolor{gray}{isocurvature} constraint, which can exclude the grey-shaded area.
The region shaded in light blue (plus the grey region) corresponds to the \textcolor{blue}{Pre-inflationary} regime, and the left-axis marks the initial misalignment $\thi$ for the axion to be DM (solid line for the quadratic regime, and dashed for the large misalignment).
In the red region, assuming that the maximum temperature is equal to 10\% of the one achieved in instantaneous reheating ($0.1\,\TRHmax$, where $\TRHmax$ is also marked on the upper axis), PQ symmetry is restored and we fall in the \textcolor{red}{Post-inflationary} regime.
The upper limits for $\HI$ and $\fa$ are due respectively to the constraints from B-modes in the CMB and black-hole superradiance.
}
\label{fig:baseline isocurvature}
\end{figure}%
This initial misalignment angle $\thi$ for the QCD axion to constitute DM is marked on the left axis in \cref{fig:baseline isocurvature}. 
The combination of \cref{eq:isocurvature bound,eq:theta_i}, together with the identification $\fainf=\fa$, defines the bound from isocurvature plotted with a grey line in the parameter space $(\HI, \fa)$ of \cref{fig:baseline isocurvature}.
The separation between the pre-inflationary regime (where $U(1)_\textsc{pq}$ is \textit{not} restored after inflation, and the misalignment angle $\thi$ in our observable Universe is set before inflation) and the post-inflationary regime (with  $U(1)_\textsc{pq}$ restored after inflation) depends on the highest temperature $\Tmax$ reached after inflation.%
\footnote{Strictly speaking, by $\Tmax$ we refer to the largest temperature reached by species that contribute in a sizeable amount to the thermal effective potential of the saxion. If the reheating of the SM and of the sector charged under $\UPQ$ occurs later, the relevant $T$ for \cref{eq:no PQ restore} can be smaller than $\Tmax$.}
The maximum possible temperature $\TRHmax$ corresponds to an instantaneous reheating, $\tfrac{\pi^2}{30} g_\star (\TRHmax)^4 = 3\He^2\MP^2$, with $\He$ the Hubble rate at the end of inflation (and generally $\He\leq\HI$), and we set $g_\star=106.75$ for definiteness.
The condition to restore $U(1)_\textsc{pq}$ after inflation and thus to be in the post-inflation regime is then%
\footnote{Some references quote $\HI/(2\pi)$ rather than $\HI$, which has a pretty small quantitative impact on \cref{eq:no PQ restore}.} 
\cite{Graham:2018jyp,Gorghetto:2018myk, Buschmann:2019icd, Redi:2022llj} 
\begin{equation}
\fa < \max\left(\Tmax, \HI
\right) \quad \text{(\textcolor{Red}{post-inflationary} regime)}
\label{eq:no PQ restore}
\end{equation}
which is of course dependent on the reheating temperature.
We show in \cref{fig:baseline isocurvature} the boundary corresponding to $\Tmax=0.1\,\TRHmax$ for reference.
The two upper boundaries on $\HI$ and $\fa$ are given respectively by the upper limit $r<0.036$ (or $\HI< 4.6 \cdot 10^{13} \GeV$) from BICEP-Planck \cite{BICEP:2021xfz} and a selection of bounds from black-hole superradiance \cite{Yoshino:2014wwa,Brito:2015oca,Brito:2017wnc,Cardoso:2018tly,Tsukada:2018mbp,Palomba:2019vxe,Kaplan:2019ako,Sun:2019mqb,Baryakhtar:2020gao}.

The four lines that we have described so far delimit a region of parameter space shaded in grey in \cref{fig:baseline isocurvature}, labelled by ``$\downarrow$ excluded by isocurvature?'', that is the subject of our paper. 
The goal of the next sections is to discuss concrete models where $\fainf > \fa$ and so the isocurvature bound is alleviated and $\UPQ$ is not restored by a non-thermal phase transition, to highlight that this region of parameter space can be accessible without significant assumptions (besides a small quartic $\l$) on the UV models for the axion and inflation.

% !TEX root = Draft_Axion-Isocurvature.tex

\section{Enhanced $f_a$ during inflation and parametric resonance}
\label{sec:par-resonance}

We assume that the strong CP problem is solved by the introduction of a complex scalar field charged under a global $\UPQ$, that we parametrise as 
\begin{equation}
\label{eq:Phi phi a}
\Phi(t,\vx) = \frac{1}{\sqrt 2} \phi(t,\vx) \exp\left(i\, \frac{a(t,\vx)}{\fa}\right)\,.
\end{equation}
We refer to the radial degree of freedom $\phi$ as the \textit{saxion} field, while the angular direction $a$ is the \textit{axion}.
The minimal Lagrangian%
\footnote{Our convention for the signature of the metric is $\text{diag}(+---)$.} 
for $\Phi$ that spontaneously breaks $\UPQ$, is
\begin{equation}
\label{eq:lagr minimal}
\mathscr L_\text{minimal}[\Phi] = |\partial_\mu \Phi |^2-
\big( \l |\Phi|^4 - \l \fa^2 |\Phi|^2 \big) + 2 \xi R |\Phi|^2 + N_\textsc{dw}\frac{a}{\fa}\frac{\as}{8\pi}G_{A\mu\nu} \widetilde G^{\mu\nu}_A,
\end{equation}
with the last term solving the strong CP problem.
Note that we have included the $R |\Phi|^2$ term because it is allowed by all symmetries, as indeed it must be since $R$ and $|\Phi|^2$ are separately allowed (as they must be or we would not have gravity or break PQ).
Choosing not to write this term is equivalent to an infinite fine-tuning of its coefficient $\xi$ to zero.
In this paper, we focus on the natural possibility that  $\xi>0$, so that the potential $V(\Phi)\supset -2\xi R |\Phi^2|$ features a negative quadratic term when $R>0$ \cite{Folkerts:2013tua, Fairbairn:2014zta}.
During inflation this term takes the value 
\begin{equation}
\label{eq:lagr R phi term}
\mathscr L_\text{minimal}[\Phi] \supset 2 \xi R |\Phi|^2  
 \overset{\text{inflation}}{\longrightarrow} 24\xi \HI^2 |\Phi|^2 \,, \qquad (\xi>0)
\end{equation}
so indeed it gives a negative contribution to the mass-squared of $\Phi$.
As we discuss in \cref{app:infl models}, other couplings are also well-motivated, such as couplings to the inflaton field or its derivative \cite{%
Bao:2022hsg%restore PQ
}
or other fields \cite{%
Jeong:2013xta,Nakayama:2015pba,Takahashi:2015waa,%coupling to Higgs in DFSZ and QCD condensation due to large Higgs vev
Ballesteros:2016xej,%SMASH
Ramazanov:2022kbd%restore PQ with Higgs-like coupling
}, and would lead to qualitatively similar dynamics to what we discuss in this section.%
\footnote{Different approaches to solve the axion isocurvature problem involve the introduction of an axion mass during inflation from PQ-breaking terms \cite{Higaki:2014ooa,Kearney:2016vqw,Co:2023mhe,Berbig:2024ufe,Carrasco:2025rud}, or the Witten effect induced by magnetic monopoles \cite{Kawasaki:2015lpf,Nomura:2015xil,Banerjee:2024ykz%monopoles and Witten effect
}.}
We will stick to the Jordan frame for simplicity for the remainder of this paper, as the physical description is independent of the chosen frame (see e.g.~\cite{Catena:2006bd,Postma:2014vaa}).
For our purposes, the choice of a specific model of inflation is irrelevant. When we discuss later the Starobinsky model, we assume that potential as a stand-in for a particular cold inflation potential. 
If one defines it in the Jordan frame, the corresponding conformal transformation might further affect the saxion potential by reducing the value of $\fainf$ with respect to $\fa$, as described in \cite{Rigouzzo:2025hza}.

The effective mass term for the saxion during inflation is then
\begin{equation}
\label{eq:fainf def}
\l \left(\fainf\right)^2 \equiv \l \fa^2 + 2 \xi R 
  \quad \longrightarrow \quad 
  \fainf = \sqrt{\fa^2 + \frac{2\xi R}{\l}} 
  \ \overset{\fainf\gg \fa}{\longrightarrow} \  \sqrt{\frac{24\xi}{\l}}  \,\HI
\end{equation}
During inflation, the saxion minimum would then lie at $\langle\phi \rangle =\fainf/\sqrt{2}$, which for $\xi>0$ lies at a greater distance than $\fa$ from the origin. 
Towards the end of inflation, $R$ decreases (both due to the decrease of $H$, and to the increasing energy fraction in relativistic species whose contribution to $R$ is loop-suppressed), and the saxion field value returns back to $\fa$.
Having a larger effective $\fainf$ during inflation naturally suppresses isocurvature perturbations as in \cref{eq:isocurvature bound}.

The idea that an enhancement of $\fa$ during inflation to a larger $\fainf$ might reduce the spread $\sth$ of the axion fluctuations in \cref{eq:sigma theta} and alleviate the isocurvature bound, dates back to \cite{Linde:1991km}, and was revived in \cite{%
Kasuya:1996ns,Kawasaki:2008jc,Folkerts:2013tua, Fairbairn:2014zta, Kearney:2016vqw%
}.
This possibility was reassessed in particular after the claim of detection of primordial B-modes in the CMB in 2014, which would have excluded the pre-inflationary option in the minimal scenario using the claimed isocurvature bounds (see \cref{fig:baseline isocurvature}).

The mechanism of an enhanced $\fainf$ to suppress isocurvature can be jeopardised by the dynamics of $\Phi$ after the end of inflation.
As the saxion field rolls towards the new minimum at $\fa \ll \fainf$, the evolution of $\phi$ might be not adiabatic, and the Hubble friction might be not sufficient to dissipate the kinetic energy of the saxion when it reaches the potential minimum.
It can then oscillate back and forth in the wine-bottle potential, possibly well above the origin.
The non-linear coupling to the axion can then give rise to a phenomenon of \textit{parametric resonance} \cite{Shtanov:1994ce,Kofman:1995fi,Kofman:1997yn}.
The oscillations of the saxion field can act as a driving term for modes of the axion with physical momentum around the Hubble rate close to the end of inflation, and drive their exponential growth.
If this phenomenon goes on for a sufficient time, the fluctuations of the angular mode can grow enough to cover the whole compact direction, forming axion strings (potentially with domain walls) and restoring non-thermally the $U(1)$ symmetry: this dynamics was analysed in general terms in \cite{Tkachev:1995md,Tkachev:1998dc,
Kasuya:1997ha,
Kasuya:1999hy% lattice study assuming RD after infl
},
and applied to the specific case of $\UPQ$ after a temporary enhancement of $\fainf$ in \cite{%
Kawasaki:2013iha,% lattice study assuming MD after infl
Harigaya:2015hha,%parametric resonance with attractor dynamics of saxion
Kearney:2016vqw,
Kobayashi:2016qld,%with N_DW=1, the strings overclose
Ema:2017krp,
Co:2017mop,Co:2020dya,%axion DM after parametric resonance
Ballesteros:2021bee%lattice simulations in SMASH
}.

Let us comment on the axion angular speed. 
In our setup, the complex $\Phi$ field sits in its inflationary vev $\fainf$ throughout the whole inflationary history, so that any initial velocity (either radial or angular) is erased.
We also do not introduce any PQ violating operator which might source a velocity $\dot a$ for the axion, as it is instead commonly assumed in the literature about the kinetic misalignment mechanism for the axion 
\cite{Co:2019wyp,Co:2019jts,%Kinetic Misalignment Mechanism
Co:2021rhi,%axion coupling to dark photon
Eroncel:2022vjg,%kinetic fragmentation
Jaeckel:2016qjp,% monodromy rather than axion
Domcke:2022wpb,Berges:2019dgr,%redistribution of charge density between rotating Affleck-Dine field and axion
Eroncel:2025qlk,Bodas:2025eca%acoustic misalignment mechanism
}.
As recently discussed in \cite{Fedderke:2025sic}, an initial angular speed can contribute to the parametric resonance and induce a non-thermal symmetry restoration for $\Phi$. This is not a concern for our scenario.

We now review the dynamics of saxion and axion in \cref{sec:par-space,app:scan}, to assess the regions of parameter space (in particular, a small $\l$) where the effect of $\xi$ in \cref{eq:fainf def}  suppresses $\sth$ and avoids isocurvature constraints, without exciting a significant parametric resonance for the axion field.

% !TEX root = Draft_Axion-Isocurvature.tex

\section{Available parameter space for pre-inflationary axion as dark matter}
\label{sec:par-space}

We begin in \cref{sec:eom} by deriving the equations of motion of saxion and axion (including the coupling $\xi R$ to gravity).  We solve them numerically in \cref{sec:scan} from the observable epoch of inflation until when the saxion settles in its final minimum at $\fa$ and give results. 
\cref{app:scan} illustrates in detail the evolution of the system and further aspects of the axion model, while \cref{app:infl models} discusses the inflationary models that we consider, together with a brief discussion of alternative couplings between the saxion and inflaton leading to a similar phenomenology.

\subsection{Equations of motion of the system}
\label{sec:eom}
We take \cref{eq:lagr minimal} as the Lagrangian density for the complex $\Phi$ field. 
The standard parametrisation of \cref{eq:Phi phi a} into radial and angular mode is not ideal to compute the evolution of the angular mode in Fourier space, because of non-linearities that couple the saxion and axion in the kinetic term (which reads in this basis $|\partial_\mu\Phi|^2 = \tfrac 12(\partial_\mu \phi)^2 + \tfrac 12 \phi^2 (\partial_\mu\theta)^2$), so that there is no perturbative limit in which these non-linear terms can be neglected.

Another possibility is to decompose the field in Cartesian coordinates \cite{Kawasaki:2013iha}, so that non-linear terms only arise from the quartic term of the potential (whose coefficient $\l$ is very small in the parameter space that we consider).
In the Cartesian parametrization,
\begin{equation}
\Phi(t,\vx) = \frac{X(t,\vx)+i Y(t,\vx)}{\sqrt 2}
\end{equation}
where we choose coordinates such that $X(t,\vx)=\fainf, Y(t,\vx)=0$ at initial times during inflation, and we can approximately identify $X$ with the saxion and $Y$ with the axion (as long as its fluctuations do not grow significantly).
Notice that this parameterisation is independent of the initial misalignment of the axion $\thi$: we are interested in exploring the growth of spatial fluctuations around the average value $\thi$, which does not affect the evolution as long as the non-perturbative QCD potential for the axion is absent.
The angular velocity $\dot Y(t,\vx)=0$ is erased by primordial inflation, and we don't assume PQ-violating operators to turn it on (see also the discussion at the end of \cref{sec:par-resonance}).
The equations of motion in this basis read (we understand the arguments $(t,\vx)$, and remind that $\fainf = \sqrt{\fa+2\xi R/\l}$)
\begin{subequations}
\label{eq:eom XY full}
\begin{align}
\ddot X +3H \dot X -\frac{1}{a^2}\nabla^2 X+\l X^3-\l \left({\fainf}^2 \textcolor{gray}{- Y^2} \right)X &= 0 \,, \label{eq:eom X full}\\
\ddot Y +3H \dot Y -\frac{1}{a^2}\nabla^2 Y \textcolor{gray}{+\l Y^3} -\l \left({\fainf}^2 - X^2\right)Y &= 0 \,. \label{eq:eom Y full}
\end{align}
\end{subequations}
We mark in \textcolor{gray}{grey} the terms that are negligible in our scenario when the parametric resonance is not significant, that is $Y\ll X,\fainf$.
From \cref{eq:eom XY full}, we can see that, as long as $\fainf$ evolves slowly, the solution for the radial mode is $X\approx \fainf$, which cancels the non-linear coupling between $X$ and $Y$ in  \cref{eq:eom Y full} for $Y$.

In order to solve numerically the equations of motion, we decompose the fields $X$ and $Y$ into background $\ovX(t),\ovY(t)$ and fluctuations $x_k(t),y_k(t)$ as
\begin{equation}
X(t,\vx)= \ovX(t) +\int \frac{\d^3x}{(2\pi)^{3/2}} x_k(t) e^{i\vk\cdot \vx} \,, \quad
Y(t,\vx)= \ovY(t) +\int \frac{\d^3x}{(2\pi)^{3/2}} y_k(t) e^{i\vk\cdot \vx} \,.
\end{equation}
The angular background velocity remains vanishing in our scenario, $\ovY(t)=0$.
As long as perturbations do not grow significantly, we can treat $x_k, y_k$ as small perturbations: $X-\ovX \ll X$, and $Y\ll X$. 
In this regime, we can write down a separate equation for the background fields $\ovX(t),\,\ovY(t)$,
\begin{subequations}
\label{eq:eom XY bkg}
\begin{align}
\ddot \ovX +3H \dot \ovX +\l \ovX^3-\l \left({\fainf}^2 \textcolor{gray}{- \ovY^2} \right) \ovX &= 0 \,, \label{eq:eom X bkg}\\
\ddot \ovY +3H \dot \ovY \textcolor{gray}{+\l \ovY^3} -\l \left({\fainf}^2 - \ovX^2\right)\ovY &= 0 \,. \label{eq:eom Y bkg}
\end{align}
\end{subequations}
In the limit of very gradual evolution of the Ricci scalar towards the end of inflation, and initial conditions $\ovX=\fainf$ and $\dot \ovY=0$, the background $\ovX$ field just tracks its minimum and \cref{eq:eom XY bkg} simplifies to
\begin{equation}
\label{eq:eom XY bkg adiabatic}
\ovX(t) \approx \fainf(t) = \sqrt{\fa^2 + (2\xi/\l) R(t)}\,,\quad \ovY\approx 0\,,
\end{equation}
with the Ricci scalar equal to $R(t)=-T^\mu_\mu/\MP^2 = (4V(\sigma)- \dot\sigma^2)/\MP^2$, where the last equality assumes that the Universe is dominated by the inflaton $\sigma$ and other relativistic species.  Of course we will not assume this adiabatic evolution of $R(t)$ since our point is to find out when the motion of $\ovX$ is too strong and produces too much parametric resonance in $Y$.%

The analytical treatment of the perturbations $x_k, y_k$ (without resorting to lattice simulations) requires to simplify the non-linear terms, dropping terms which are perturbatively small as long as $Y\ll X$ and $X-\ovX \ll X$. 
Concretely, by denoting schematically $X=\ovX +\int_k x_k$, $Y =\int_k y_k$, we can neglect the grey terms in
\begin{equation}
\begin{aligned}
\l X^3 & = \l \ovX^3+3\l \ovX^2 \int_k x_k 
  \textcolor{gray}{+3\l \ovX \int_k\int_q x_k x_q + \l \int_k \int_q \int_p x_k x_q x_p} \\
\l X^2Y & = \l \ovX^2\ovY +\l \ovX^2 \int_k y_k 
  \textcolor{gray}{+2\l \ovX \int_k\int_q x_k y_q + \l \int_k \int_q \int_p x_k x_q y_p}
\end{aligned}
\end{equation}
to obtain the linear equations of motion for the perturbations (which are valid as long as $Y\ll X$ and $X-\ovX \ll X$) 
\begin{subequations}
\label{eq:eom XY pert}
\begin{align}
\ddot x_k +3H \dot x_k + \left(\frac{k^2}{a^2} + \l \left(3\ovX^2 - {\fainf}^2\right) \right) x_k &= 0 \,, \label{eq:eom X pert}\\
\ddot y_k +3H \dot y_k + \left(\frac{k^2}{a^2} + \l \left(\ovX^2 - {\fainf}^2\right) \right) y_k &= 0 \,. \label{eq:eom Y pert}
\end{align}
\end{subequations}
\cref{eq:eom Y pert} nicely highlights the term responsible for the growth of perturbations via parametric (or tachyonic) resonance, which is only negligible if the evolution of the radial mode is almost adiabatic, $\ovX(t) \approx \fainf(t)$ as in \cref{eq:eom XY bkg adiabatic}.
In the regime when the growth of the angular direction (approximately coinciding with $Y$) is under control, then the fluctuations of the radial direction in \cref{eq:eom X pert} remain small as well.

\subsection{Numerical solution across the parameter space}
\label{sec:scan}
We solve numerically \cref{eq:eom X bkg,eq:eom Y pert} in the background evolution of the inflaton during the observable epoch of inflation and the preheating phase. 
For our illustrative purposes, we consider two inflationary models (described in detail in \cref{app:infl models}). 

The first one is a model of \textit{warm inflation}, where a thermal bath is maintained during inflation by the dissipation of the inflaton kinetic energy. 
The Ricci scalar smoothly vanishes as radiation takes over at the end of inflation.
The second one is the Starobinsky model of \textit{cold inflation}, with a preheating phase at the end inflation when the inflaton $\sigma$ rolls around its minimum. During this matter-dominated phase, the Ricci scalar oscillates as $R(t)= H(t)^2(3+9\cos(2\ms t)) \cdot \rho_\sigma(t)/\rho_\mathrm{tot}(t)$. 
We model reheating as a decay of the inflaton into relativistic species via a decay rate for which we take three example values: $\Gs = 10^{-(1,3,5)}\ms$.

In the main text we will give the results for the warm inflation model.  This warm inflation model yields a slightly larger region of parameter space which is safe from isocurvature compared to the particular cold inflation model we used.  This is because our chosen warm inflation model ends more smoothly, so the saxion relaxes towards its ultimate late-time minimum more gradually than in the cold inflation model we chose, where the end is more abrupt.  The more abrupt end will tend to drive somewhat more parametric resonance production of the axion.  The results for the particular cold inflation model we chose are presented in \cref{app:infl models} (\cref{fig:cold par space}). We have not done a scan over all possible inflation models (nor over other possible UV realizations of the axion/saxion).  We just chose two inflation models to illustrate the model-dependence.  So in particular we certainly cannot state a theorem that warm inflation will in general give a larger parameter space than cold inflation.

We treat the inflationary Hubble rate $\HI$ (conventionally defined at 60 $e$-folds before the end of inflation) as a free parameter for the background evolution of the scale factor and $R$ in the two models.  
We sample also the three free parameters $(\fa, \xi,\l)$ for the axion Lagrangian, for a total of four free parameters. 
We impose the following conditions on the parameter space.

\begin{figure}[h!]%\centering
\hspace{7em}\includegraphics[height=15em, trim = 0 0 293 0, clip]{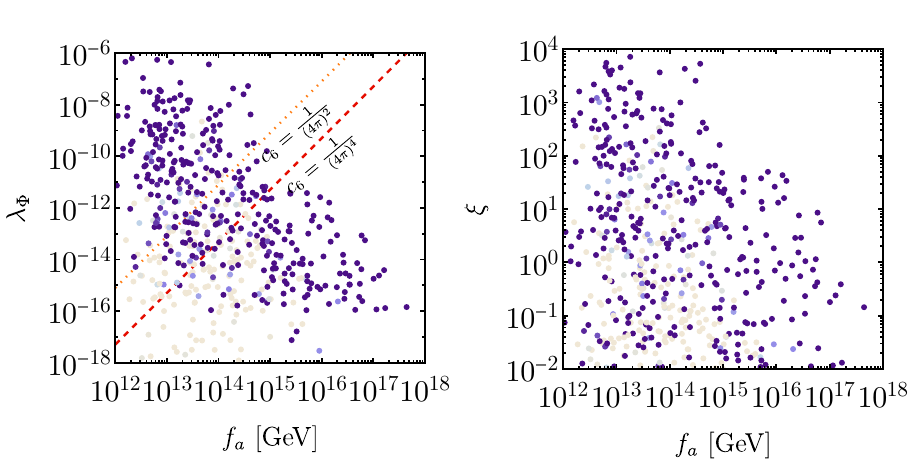}\vspace{-1em}\\
\hspace*{7em}\includegraphics[height=15em, trim = 440 0 0 0, clip]{1_warm_2_fa-xi-l}
\caption{Points of our scan for the model of Warm Inflation, projected on the 2-dimensional subspaces of $(\fa,\l,\xi)$. 
The colour code of each point corresponds to the resulting $\dthpr$.
The dotted and dashed red lines in the first plot mark the lower boundaries where the points of the scan are not affected by the presence of a higher-order term $\frac{c_6}{\MP^2}|\Phi|^6$ in the saxion potential.}
\label{fig:points scan warm}
\end{figure}

\begin{itemize}
\item We focus on the \textit{pre-inflationary} parameter space, requiring then from \cref{eq:no PQ restore} that $\fa > \Tmax$. For each inflationary model, the ratio between $\Tmax$ and the temperature corresponding to an instantaneous reheating $\TRHmax \equiv (90 \HI^2 \MP^2/(\pi^2\gs))^{1/4}$ is fixed, so  
\begin{equation}
\label{eq:Tmax}
\Tmax = 8.4\cdot 10^{13} \GeV \parfrac{\HI}{10^{10}\GeV}^{1/2} \cdot \epsilon_i \,, \quad \epsilon_i = 0.03 - 0.32 \text{ for our infl.\,models}.
\end{equation}
\item The term $2\xi R|\Phi|^2$ in \cref{eq:lagr minimal} corresponds to a negative contribution to the \textit{total energy density during inflation}, which should not overcome the inflaton one. This amounts to 
\begin{equation}
\label{eq:no-ads}
\left| V\left(\langle \Phi\rangle = \tfrac{1}{\sqrt{2}}\fainf\right)\right| \approx \frac{\xi^2(12\HI^2)^2}{\l}  <3\HI^2 \MP^2\,.
\end{equation}
\item The coefficient $\xi$ can be bounded from above by \textit{unitarity} arguments. In the scattering $\Phi\Phi\to \Phi\Phi$ mediated by a graviton exchange, the operator $2\xi R|\Phi|^2$ contributes with a series of couplings of $\Phi\Phi$ to gravitons. 
The leading term comes at mass dimension 5 \cite{Burgess:2009ea, Barbon:2009ya}, with an amplitude of order $\xi^2 E^2/\MP^2$, where $E$ is the energy of the process. 
We impose that this amplitude is smaller than 1 at the highest energy scale of the problem, the inflationary scale $(\HI\MP)^{1/2}$.
\item The smallness of the quartic coupling $\l$ that we consider means we should check the size of quantum corrections (see also \cite{Markkanen:2018bfx,Kozow:2022whq}). 
In the model described by \cref{eq:lagr minimal}, the argument of dimensional analysis (by restoring $\hbar$ to make couplings dimensionful \cite{Panico:2015jxa}) shows that the only terms which might correct the quartic coupling at loop order $\hbar/(4\pi)^2$ are $\propto \xi R$ or $\propto \mphi^2 (= \l \fa^2)$. Any correction to $\l$ would then be suppressed by the second power of one of them over $\fa^4$ or $\MP^4$. All of these combinations are small enough in our sample.
Or of course the loop correction could  be proportional to $\l$ itself, but then is necessarily small enough.
We comment in \cref{app:friction} on an analogous requirement in case the saxion couples \textit{à la} KSVZ via a Yukawa coupling to coloured fermions.
To be thorough, we should also estimate gravitational loop corrections to $\l$.  Some of these graviton loops may be quartically divergent.  The maximum contribution these can make is $\sim \frac{1}{(4 \pi)^2} \frac{\text{cutoff}^4}{\MP^4}$.  To be conservative, if we take the cutoff of the gravitational EFT to be as high as $\fa$ we still find that this contribution is negligible in most of parameter space and would require a mild fine-tuning of $\l$ for the points in the lower-right corner of \cref{fig:points scan warm}.  Even this may be an overestimate since whatever UV physics cuts off these graviton loops, for example supersymmetry, could come in at a scale well below $\fa$.  Thus it seems likely that gravitational corrections do not give too large a contribution to $\l$ and do not require fine-tuning.
\item An upper limit in the parameter space $(\HI,\fa)$ can be obtained by requiring the validity of the EFT defined by the gravitational terms $\frac 12\MP^2 R + 2\xi |\Phi|^2 R$ for $\langle\Phi\rangle =\fainf \approx \sqrt{24 \xi/\l}\HI$, requiring that the Einstein-Hilbert action is the leading term in an expansion in powers of $\xi |\Phi|^2/\MP^2$.
\end{itemize}
We perform a scan of the 4-dimensional parameter space $(\HI, \fa, \xi,\l)$ as follows.
We draw a random sample of points satisfying the isocurvature bound \cref{eq:isocurvature bound} with $\fainf$ defined as in \cref{eq:fainf def}, and further impose the constraints described in the previous bullet points. 
The conditions that we impose on the sample are collected in Eqs.~(\ref{eq:no PQ restore rough})-(\ref{eq:unitarity lambda}).\\
We compute numerically the evolution of the inflationary background over the last 60 $e$-folds of inflation and the subsequent reheating, with $\HI$ as a free parameter (we neglect the small backreaction of $2\xi R|\Phi|^2$ on the inflaton). In this section, we show the results for a model of warm inflation, which is described in \cref{app:infl models} together with a comparison with a model of cold inflation.\\
Then, we compute the evolution of the saxion background $\ovX(t)$ in this background, using \cref{eq:eom X bkg}. The initial condition for $\ovX$ is not important, as it quickly relaxes to its minimum during inflation.\\
Finally, we solve the linearised equations of motion for the perturbations $y_k$ (Eq.~\ref{eq:eom Y pert}) for a range of axion modes with physical momentum $k\sim \mathcal O(0.1-10) \kend$, with $\kend =\ae \He$ crossing the Hubble radius at the end of inflation. The initial condition for each mode $y_k$ is set by the quantum fluctuations of the Bunch-Davies vacuum, which at the time $t_{\cross}$ when the mode crosses the Hubble radius reads $|y_k(t_{\cross})| \approx H_{\cross}/\sqrt{2k^3}$.
We perform this numerical calculation for enough time to detect whether the parametric resonance induced on the axion is strong enough to amplify its power spectrum more than $(\pi \fa)^2$ and restore $\UPQ$. We compute the power spectrum (at a late time when $\ovX(t)\approx \fa$ and $Y(x)\approx a(x)$) as $\langle (a(t,x)-a_i)^2\rangle \approx \langle Y(t,x)^2\rangle =\int |y_k(t)|^2 \d^3 k$. 

\begin{figure}\centering
\includegraphics[width=.85\textwidth]{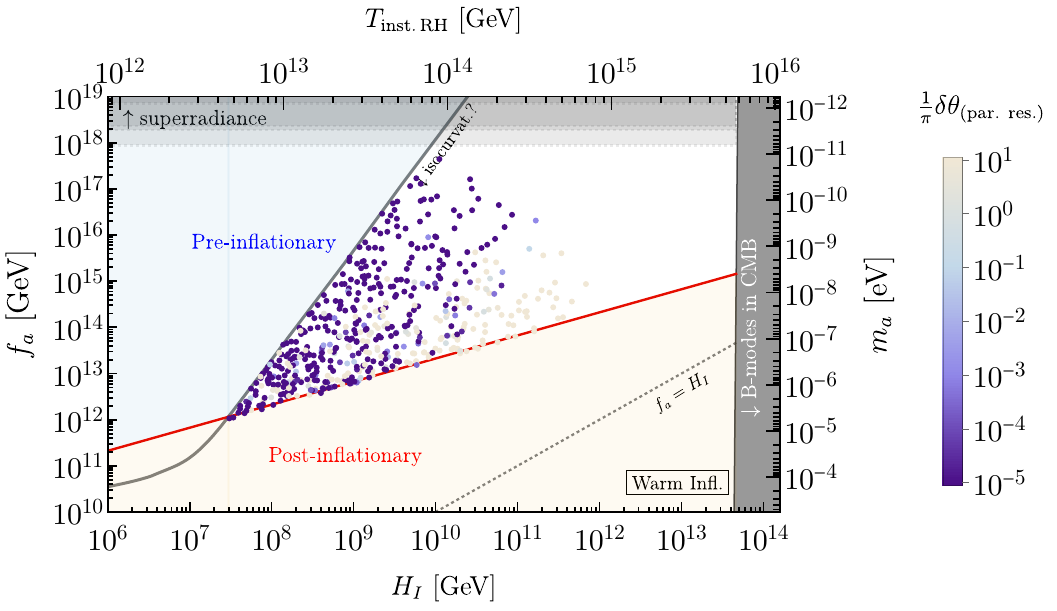}
\caption{Same as \cref{fig:baseline isocurvature}, showing the amplification of axion fluctuations for all the points of our scan in $(\HI,\fa,\xi,\l)$ for our model of Warm Inflation. 
The color of each point indicates how much parametric resonance drives the axion modes, as shown in the legend on the right.
Blue and light blue points do not feature a strong parametric resonance, and successfully avoid the isocurvature constraint without overproducing axions. 
For light-orange points, axion strings are formed due to parametric resonance.}
\label{fig:warm scan}
\end{figure}

The result of this scan of the parameter space is shown in \cref{fig:warm scan} as projected on the $(\HI,\fa)$ plane.
The colour code marks the enhancement of axion fluctuations as a result of the dynamics after the end of inflation. 
Points in blue experience little or no parametric resonance, whereas the evolution of light blue and rose points amplifies the fluctuations in the misalignment angle to the point where our linearised analysis breaks down and $\UPQ$ is non-thermally restored, leading to the overproduction of axions at late times (in this part of parameter space). 
More details about these results are given in \cref{app:par res} and \cref{fig:warm big par-res,fig:warm small par-res}.

\begin{figure}\centering
\includegraphics[width=.6\textwidth]{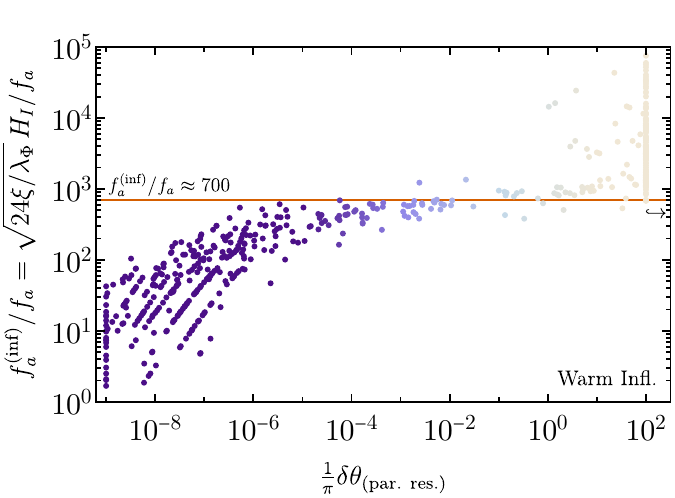}
\caption{Growth of the axion fluctuations due to parametric resonance as a function of the enhancement of $\fa$ during inflation, $\fainf\approx\sqrt{24 \xi/\l} \HI$, for our scenario of warm inflation.
The red horizontal line marks $\fainf/\fa\approx 700$, which roughly divides the points featuring inefficient or short parametric resonance (in blue) and string formation (light orange).
Points which display an amplification $ \tfrac{1}{\pi}\dthpr <10^{-9}$, or $>10^{2}$, are all marked on the respective threshold. 
Some points on the lower left align on diagonal lines because of our sampling of integer powers of $\xi/\l$.}
\label{fig:warm fainf-amplPS}
\end{figure}
In order to gain some analytical insight into the scatter plot of \cref{fig:warm scan}, it is very useful to examine for every point in the scan the amplification of the power spectrum under parametric resonance:
\begin{equation}
\label{eq:dthpr def}
\frac 1\pi \dthpr \equiv \frac{\langle a(x)^2\rangle^{1/2}}{\pi \fa}
\end{equation}
versus the enhancement of $\fa$ during inflation $\fainf/\fa \approx \sqrt{24\xi/\l} \HI/\fa$.
The result in \cref{fig:warm fainf-amplPS} shows that the enhancement $\fainf/\fa$ is a very good proxy to predict the strength of the subsequent parametric resonance.
The origin of this correlation is clear, as the linearised \cref{eq:eom Y pert} for the perturbations $y_k$ has a non-linear driving source proportional to the displacement of $\ovX(t)$ from the minimum $\fainf(t)$. 
The larger is the excursion of $\fainf$ at the end of inflation when moving to $\fa$, the longer is the duration of the parametric resonance for $y_k$. \cref{app:par res} discusses this point with further details.

As a result, we can posit an approximate condition to ensure that $\UPQ$ is not restored due to parametric resonance: for the model of warm inflation considered in this section, this condition is $\fainf/\fa \lesssim 700$. 
As clear in \cref{fig:warm fainf-amplPS}, when this condition is satisfied, the amplification $\dthpr$ of the power spectrum of the misalignment angle remains $< \mathcal O(10^{-3})$, ensuring the validity of our linearised analysis.
This condition is model dependent: the details of the saxion-inflaton coupling and of the model for inflation and reheating, affect this quantitative threshold. Depending on the complexity of the dynamics, and on the possibility (for cold inflation) that the oscillations of the inflaton at preheating might act as additional source of parametric resonance, this condition to prevent string formation might be a bit less sharp than in \cref{fig:warm fainf-amplPS}.
At any rate, also for the models of cold inflation that we consider in \cref{app:infl models} we find that the condition $\fainf/\fa \lesssim \mathcal O(100)$ is an excellent proxy to avoid the formation of axion strings.

We can now summarise for completeness all the constraints on the 4-dimensional parameter space $(\HI,\fa,\xi,\l)$ discussed in this section. To ease the reader, we mark in \textcolor{constant}{grey} the constant coefficients.%
\begin{subequations}
\begin{align}
\textcolor{constant}{\parfrac{90\MP^2}{\pi^2\gs}^{1/4}\epsilon_i} &< \frac{\fa}{\HI^{1/2}}
&&\text{
avoidance of PQ restoration at reheating, \cref{eq:Tmax}
} \label{eq:no PQ restore rough}\\
\textcolor{constant}{\frac{2.0\cdot 10^{-6}}{\GeV^{7/5}}} &> \frac{\fa}{\HI^{12/5}}
&&\text{
standard isocurvature for axion DM, \cref{eq:isocurvature bound,eq:theta_i}
} \label{eq:isocurvature naif}\\
\textcolor{constant}{\frac{\GeV^{7/6}}{7.5\cdot 10^{-7}}}\frac{\xi}{\l} &> \fa^{7/6}
&&\text{
isocurvature bound for axion DM with $\fainf$, \cref{eq:fainf def}
} \label{eq:isocurvature axion DM}\\
 \textcolor{constant}{\frac{4\sqrt{3}}{\MP}}  \frac{\xi}{\l^{1/2}} &< \frac{1}{\HI}
&&\text{
saxion energy density not exceeding $\rho_\text{tot}$, \cref{eq:no-ads}
} \label{eq:no AdS}\\
\frac{1}{\textcolor{constant}{\MP^{1/2}}}\xi &< \frac{1}{\HI^{1/2}}
&&\text{
unitarity bound on $\xi$ from graviton-mediated $\Phi \Phi \to \Phi\Phi$
} \label{eq:unitarity xi}\\
\textcolor{constant}{\frac{4\sqrt{6}}{\MP}}\frac{\xi}{\l^{1/2}}
&
\lesssim \frac{1}{\HI}
&&
\text{EFT validity ($2\xi (\fainf)^2R$ sub-leading wrt $\tfrac 12\MP^2R$)} \label{eq:unitarity lambda} \\
\textcolor{constant}{\mathcal O\left(10^{-2}\right)}\sqrt \frac{\xi}{\l} &\lesssim \frac{\fa}{\HI}
&&\text{
avoiding string formation
(\cref{fig:warm fainf-amplPS})
} \label{eq:no strings}
\end{align}
\end{subequations}
All of these conditions are set analytically except the last condition \cref{eq:no strings} which is an approximate result of our numerical analysis.  The numerical coefficient of the last condition would be $\approx \sqrt{24}/700$ in the warm inflation model we consider but is different in other models, though in general is $\sim 10^{-2}$.

\begin{figure}[h]\centering
\includegraphics[width=.8\textwidth]{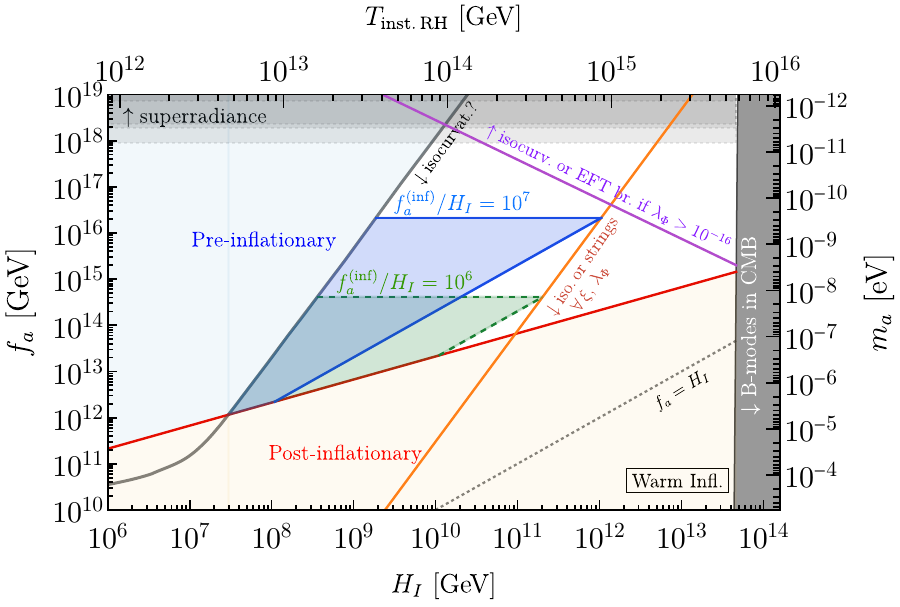}
\caption{The blue and green regions show the parameter space for fixed values of $\fainf/\HI\approx \sqrt{24 \xi/\l}$ where the QCD axion can be the DM and not violate the isocurvature constraint (with the  $2\xi R|\Phi|^2$ coupling and a small quartic coupling $\l$).
The plot shows the regions, in the parameter space typically subject to isocurvature, where $\fainf >\fa$ suppresses isocurvature without sourcing a parametric resonance for the axion, for the model of Warm inflation. 
In general the whole region that is rescued from the isocurvature constraints is the one enclosed by the purple and orange (as well as grey and red) lines.
}
\label{fig:warm xi-over-lambda}
\end{figure}
Being equipped with \cref{eq:no strings} to determine the formation of axion strings, we can understand analytically the results of the scatter plot in \cref{fig:warm scan}. 
Let us fix the ratio $\xi/\l$, or equivalently $\fainf/\HI$. 
For a fixed value there is an allowed wedge in the plane $(\HI,\fa)$, as marked in \cref{fig:warm xi-over-lambda} for two choices of $\fainf/\HI$ (the blue and green regions).
The isocurvature bound with $\fainf$, \cref{eq:isocurvature axion DM}, gives the horizontal upper limit of these regions.
The condition \cref{eq:no strings} to prevent string formation translates into a diagonal upper limit on $\HI$ as seen in the right hand sides of the blue and green regions in \cref{fig:warm xi-over-lambda}. 
As we vary $\fainf/\HI$, the envelope of these wedges identifies the boundary marked in orange in \cref{fig:warm xi-over-lambda}. If we denote by $R_{\fa}$ the upper limit on $\fainf/\fa$ identified in \cref{fig:warm fainf-amplPS} (in that case, $R_{\fa}=700$), this right-hand side boundary of the parameter space ``rescued'' from isocurvature reads (using \cref{eq:isocurvature axion DM} for the isocurvature bound)
\begin{equation}
\fa> 8.0 \cdot 10^{15}\GeV \parfrac{\HI}{10^{11}\GeV}^{12/5}
   \parfrac{R_{\fa}}{100}^{-12/5} \quad \text{(isocurv.\;and strings can be avoided)}
\label{eq:rescued}
\end{equation}
This relation is based on the quadratic approximation for the axion potential; for $\fa<10^{12}\GeV$, the orange line would remain parallel to the isocurvature line shown in grey in Fig.~\ref{fig:warm xi-over-lambda}.
Finally, there is an upper limit on $\fa$ that can be derived by the combination of \cref{eq:unitarity lambda,eq:isocurvature axion DM}. This bound reads
\begin{equation}
\fa< 0.16\,\MP \parfrac{\HI}{10^{11}\GeV}^{-6/7}
   \parfrac{\l}{10^{-16}}^{-3/7} \quad \text{(isocurv.\;and EFT break.\;can be avoided)}
\label{eq:rescued upper}
\end{equation}
and is marked in purple in \cref{fig:warm xi-over-lambda}, with a choice of $\l = 10^{-16}$ for the figures of the paper (a benchmark choice which selects most of our blue points in \cref{fig:points scan warm}).
This line also roughly coincides with an anaologous combination of the isocurvature constraint \cref{eq:isocurvature axion DM} and the requirement that the saxion VEV $\fainf$ during inflation does not reach Planckian values.

\begin{figure}[h]\centering
\includegraphics[width=.8\textwidth]{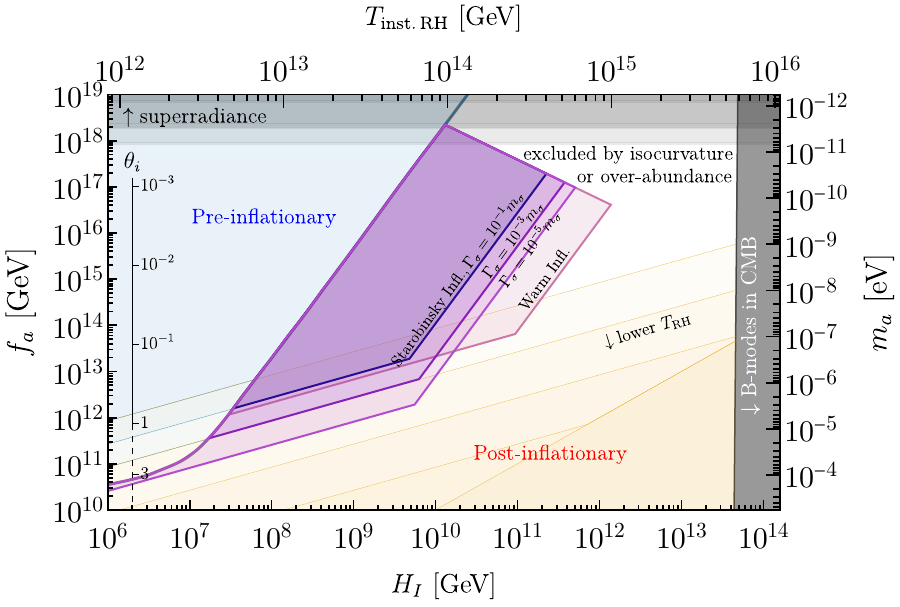}
\caption{Parameter space in the plane $(\HI,\fa)$ where the QCD axion can be the DM respecting the isocurvature constraint, in presence of the $\xi R|\Phi|^2$ coupling and a small quartic coupling $\l$.
The plot shows the regions where an enhancement of $\fainf$ suppresses isocurvature without exciting the axion via parametric resonance, for the inflationary models considered in this paper. 
The purple regions are not ruled out by isocurvature constraints despite the general lore, as we have shown in this paper. 
}
\label{fig:isocurvature-rescued}
\end{figure}
The boundaries identifying the regions of pre-inflationary axion ``rescued'' from the isocurvature constraint have a moderate model dependence, for which we show a sample in \cref{fig:isocurvature-rescued}.
The bottom boundary of the region has the usual dependence on the maximum temperature $\Tmax$ attained (by the bath of the SM and PQ sector) after inflation.
The right-hand side boundary depends instead on how easy it is to drive the parametric resonance for the axion. 
In this plot we show the results for a warm  and cold inflation model (whose details are in \cref{app:infl models}). 
Warm inflation displays a large gradual decrease of $H(t)$ over the last 60 $e$-folds of inflation and the exponential decay of $R$ (and inflaton value) at reheating: both aspects shorten the duration of parametric resonance, making it ineffective in a wide region of parameter space.
For cold inflation, the decrease of $H(t)$ during inflation is usually more abrupt towards $t_e$, and $R$ then oscillates at a frequency $2\ms$, potentially inducing an additional driving term for the axion evolution. 
This explains the dependence on the duration of reheating and the inflaton decay rate $\Gs$. 

We plot in \cref{fig:points scan warm} the points of our scan for the model of Warm Inflation projected on the 2-dimensional subspaces of the variables $(\fa,\l,\xi)$ (while \cref{fig:warm scan} shows them in the $(\HI,\fa)$ plane). 
Analogously to that figure, the colour code of the points marks in blue the points where $\dthpr \lesssim \mathcal O(10^{-3})$ and strings are not formed.

The top left panel of \cref{fig:points scan warm} also shows the lines below which a higher-order operator $\frac{c_6}{\MP^2}|\Phi|^6$ dominates the potential at $|\Phi| < \fa$, for two values of $c_6$ suggestive of a 1-loop ($c_6=\tfrac{1}{(4\pi)^2}$) or 2-loop ($c_6=\tfrac{1}{(4\pi)^4}$) origin.
While we show these lines for completeness and to highlight the mild impact of 6th-order terms on our sample points, it is very plausible that a model providing a small quartic implies  automatically also small higher-order corrections.

Let us add some comments about further effects that we consider in connection to this dynamics.

\begin{itemize}
\item We neglect the fluctuations $x_k$ of the saxion, which display a similar growth as the axion fluctuations and are then irrelevant for us, as long as $\tfrac 1\pi \dthpr$ is safely $\ll 1$.
One should consider also that the saxion, while passing over the wine bottle potential across the origin, might stop in the angle $\theta_i$ (as in the top-left panel of \ref{fig:warm small par-res}) or in $\theta_i+\pi$ (as in \cref{fig:warm big par-res}).
This only affects the initial value of the misalignment angle, which we assume to be the right one to yield the axion as DM.
We do not need to worry instead that the final angular position ($\thi$ or $\thi+\pi$) might vary across different spatial patches, depending on the local quantum fluctuations of the saxion. 
The reason is that, for the ``rescued'' region of \cref{fig:isocurvature-rescued}, the saxion crosses the origin less than 10 times, and in that region the relative quantum fluctuations of the saxion $(\HI/(2\pi))/\fa$ are much smaller than $1/10$, implying that the saxion stops in the same field position across the whole observable universe.

\item The condition that we imposed in \cref{fig:warm fainf-amplPS}, $\dthpr \lesssim \mathcal O(10^{-3})$, might need be strengthened if the background misalignment angle $\pi-\thi$ were very small. 
The reason is that spatial fluctations  of the axion (due to $\dthpr$) would lead to the formation of some axion strings in our observable universe, and change the prediction for the DM abundance.
This point only concerns $\fa \lesssim 10^{10}\GeV$, and it implies that we can still ``rescue'' the pre-inflationary axion from isocurvature even in the large-misalignment region, at the cost of a larger value of $\xi/\l$.

\item When the saxion settles in the minimum, its oscillations redshift as matter, and their energy density grows relative to the ambient relativistic energy density. 
These saxion excitations eventually decay, either to the SM or axions (or other species decoupled from the SM).\\
The decay of the saxion depends on the content of the PQ sector. 
A model-independent channel is the decay to axions, proportional to $\mphi^3/\langle\phi\rangle^2 \sim \l^{3/2} \fa$ \cite{Eroncel:2024rpe}. 
Another channel that is generically present is the decay to SM gluons, also proportional to $\mphi^3/\langle\phi\rangle^2$. 
Other decay channels, as e.g.~to KSVZ fermions (when kinematically allowed, see \cref{app:friction}), can deplete the saxion energy density into the SM plasma.
For the ``safe'' blue points of the scan shown in \cref{fig:warm scan}, we find that the decays to axions or gluons happen around a similar temperature (due to the same parametric scaling of their decay rates) around $T\sim 10^{2-9}\GeV$.\\
For about half of the points, the decay occurs after the domination by the saxion of the energy budget of the Universe.
In this case, we might worry about isocurvature perturbations on CMB scales. Saxion perturbations on large scales are tiny at early times (the saxion being a massive degree of freedom during inflation \cite{Linde:1990flp,Kolb:2023ydq,Racco:2024aac}). 
They inherit the adiabatic perturbations on large scales from the spatial dependence of the Hubble rate: there is effectively a single clock for curvature perturbations.
We find then no general issue with the saxion inducing a matter-dominated epoch, given that it decays way before BBN in the parameter space of interest to us.\\
If the saxion dominantly decays to axions we find that, for a large fraction of the points (80\% for the scan in \cref{fig:warm scan}), the axion number density of hot axions injected by the saxion decay is smaller than the number density of cold axions produced from misalignment after the QCD transition. The points which do not satisfy this condition lead to too hot DM.\\
In summary, a detailed reconstruction of the dynamics and decay of the saxion energy density (as e.g.~in \cite{Moroi:2013tea,Moroi:2014mqa,Gouttenoire:2021jhk, Eroncel:2024rpe}) can clarify the cosmological evolution and, up to some model dependence, might identify some regions of parameter space that are not allowed.
We find that the issue of saxion decay does not pose a problem over most of the points of our scan, and does not generically challenge the possibility to avoid isocurvature via an enhancement of $\fainf$.

\end{itemize}

% !TEX root = Draft_Axion-Isocurvature.tex

\section{Conclusions}
\label{sec:conclusions}

It has generally been claimed that there are strong isocurvature constraints on the minimal QCD axion model at high $f_a$ and high inflationary Hubble scale $\HI$, see \cref{fig:baseline isocurvature}.
We have shown that in fact much of this region is not ruled out.  \cref{fig:isocurvature-rescued} shows in purple the regions of the minimal QCD axion that are allowed, contrary to the normal claim about isocurvature.

The isocurvature bounds are highly UV dependent.  While it is possible for the region shown in  \cref{fig:baseline isocurvature} to overproduce isocurvature perturbations, it does not have to happen.  Whether or not a particular point is ruled out depends on UV parameters of the PQ-breaking potential as well as the models of inflation and reheating.
The minimal model for the PQ-breaking potential has the form shown in \cref{eq:lagr minimal}.
When $\l$ is small and $\xi \sim \mathcal{O}(1)$ (or really the ratio $\l / \xi \ll 1$) then during inflation the axion will have a much larger effective PQ-breaking scale $\fainf$ than it will in the later universe.  This automatically suppresses the isocurvature perturbations.

While this was realized previously, it was believed that this possibility was ruled out because the large drop in the saxion field $\phi$ at the end of inflation would drive parametric resonance of the axion, creating axion strings and thus overproducing axion dark matter over most of this region.
We have shown that in fact this does not necessarily happen.  There are many points in the UV parameter space of this model where parametric resonance is weak and the axion is not overproduced.
The strength of this parametric resonance depends both on the UV parameters of the minimal QCD axion model in \cref{eq:lagr minimal} as well as the model of inflation and reheating.  We chose two regular inflation models and demonstrated that many parts of the parameter space work in each model.  The exact region in the $\fa$-$\HI$ plane which is rescued from isocurvature this way depends on the models of inflation and rehehating as shown in \cref{fig:isocurvature-rescued}.

\begin{figure}[t]\centering
\includegraphics[width=.8\textwidth]{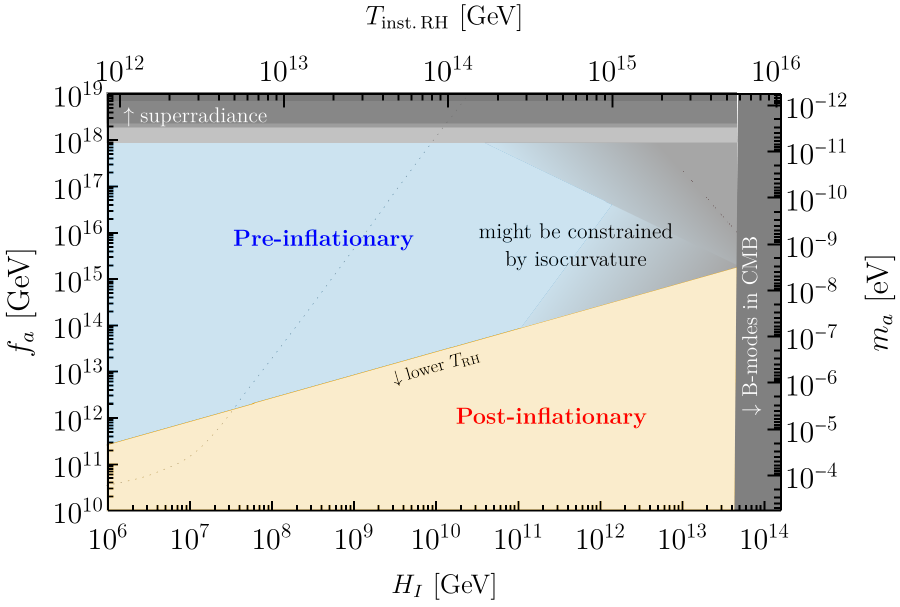}
\caption{Summary plot of the region (shaded in grey) that is  not rescued from the isocurvature constraint even by an enhancement of $\fainf$ during inflation in the inflation models we chose. 
This plot assumes a threshold $\fainf/\fa<R_{\fa}$ (with $R_{\fa}=700$) for which parametric resonance of the axion can be avoided, as given in \cref{eq:rescued}.
Note that a different inflation model or a different UV model of the axion might cause even the grey shaded regions to be allowed and thus we cannot say conclusively that they are ruled out by isocurvature.
The reheating line (in orange) assumes $\Tmax=\TRHmax/\sqrt{10}$.}
\label{fig:isocurvature-rescued-summary}
\end{figure}

The strength of the parametric resonance is affected by the speed with which $R$ drops around the end of inflation and reheating.  If $R$ and hence $\fa$ change rapidly, then the transition is too sudden and a strong parametric resonance is driven with large axion production.  However if $R$ and $\fa$ change more slowly, the minimum of the saxion potential moves more adiabatically and the parametric resonance is weak.
Whether the drop in $\fa$ from the start of inflation until late times is adabiatic or not depends on the model of inflation and reheating as well as UV parameters of the saxion potential.  We have shown that it is easy to find points in UV parameter space where the parametric resonance is weak and isocurvature is suppressed without overproduction of the axion\footnote{See for example the blue points in \cref{fig:warm scan}.}.

Note the parts of \cref{fig:isocurvature-rescued-summary} which are shaded in grey are not allowed (by isocurvature or overclosure)  with the specific inflation/reheating models we consider.  Other models of course might extend to more of this parameter space.  But interestingly even in the models we consider there is a change to the physics in those regions.  They are not allowed because, while the isocurvature production can be suppressed, in that case the parametric resonance is too strong and we enter an effectively post-inflationary scenario where the axion abundance is too large at those high $\fa$.  This is why those regions are labelled in \cref{fig:isocurvature-rescued} with ``excluded by isocurvature or over-abundance''.  But that does suggest that this region may be rescued by specific models along the lines of those proposed to explain the small $\thi$ (e.g.~an entropy dump in the early universe).  We have focused only on the minimal QCD axion model so have not pursued this avenue, but just note that the effects of the $R |\Phi|^2$ term change the physics in these regions as well.

We focused specifically on the effects of the $R |\Phi|^2$ term as it cannot be forbidden.  However $\Phi$ can also generically have other couplings, for example to the inflaton, which can similarly create a large $\Phi$ mass during inflation and would likely also remove significant parts of the isocurvature bounds (see \cref{app:infl models}). 
It would be interesting to work through the same exercises for those couplings to see what regions of the QCD axion parameter space are additionally allowed.
More generally, we focused on a minimal 4D field-theoretic UV completion for the QCD axion just for simplicity.  However it would also be interesting to consider extra-dimensional or string-inspired UV models for the axion.

The issue of isocurvature constraints (and the coupling between cosmology and the QCD axion in general) is important because many new measurements are coming, both in cosmology and axion detection.  A large $\fa$ near fundamental scales such as the GUT, string, or Planck scales is highly theoretically motivated.  This significant weakening of the isocurvature constraints means that there is now more room for the QCD axion to have such a large $\fa$ even with a high Hubble scale for inflation.

\acknowledgments
%\begin{acknowledgments}
We thank Marco Gorghetto, Ed Hardy, Junwu Huang, Liam McAllister, Riccardo Natale, Maximilian Ruhdorfer, Ben Safdi, Géraldine Servant and Sebastian Zell for helpful discussions.

\noindent
The authors acknowledge support by NSF Grant PHY-2310429, Simons Investigator Award No.~824870, DOE HEP QuantISED award \#100495, the Gordon and Betty Moore Foundation Grant GBMF7946.
D.R.~is supported at U.~of Zurich by the UZH Postdoc Grant 2023 Nr.\,FK-23-130.
%\end{acknowledgments}

%%%%%%%%%%%%%%%%%%%%%%%%%%%%%%%%%%%%%%%%%%

\appendix

\section*{Appendices}

% !TEX root = Draft_Axion-Isocurvature.tex
\section{Details about the evolution of saxion and axion}
\label{app:scan}

In this appendix, we provide more details about the linearised numerical evolution of the saxion-axion that we perform (\cref{app:par res}), and we discuss model-dependent features of the PQ model that can enrich the dissipative dynamics, and possibly assist in weakening the parametric resonance (\cref{app:friction}).

\subsection{Examples of parametric resonance for the axion}
\label{app:par res}
In this appendix, we expand the discussion in \cref{sec:scan} around \cref{fig:warm scan} of the coupled dynamics of saxion and axion leading to parametric resonance for the axion.

During inflation, the saxion lies in the vicinity of the minimum $\fainf(t)$, trailing behind it as $R$ evolves in time through $\HI(t)$.
The time derivative of the distance $\ovX(t)-\fainf(t)$ scales roughly as $\dot \HI(t)\sim \HI(t)^2$ during inflation, and its evolution at the end of inflation depends more sensitively on the inflationary potential closer to its minimum, and on the reheating dynamics.

\begin{figure}[th]\centering
\includegraphics[width=\textwidth]{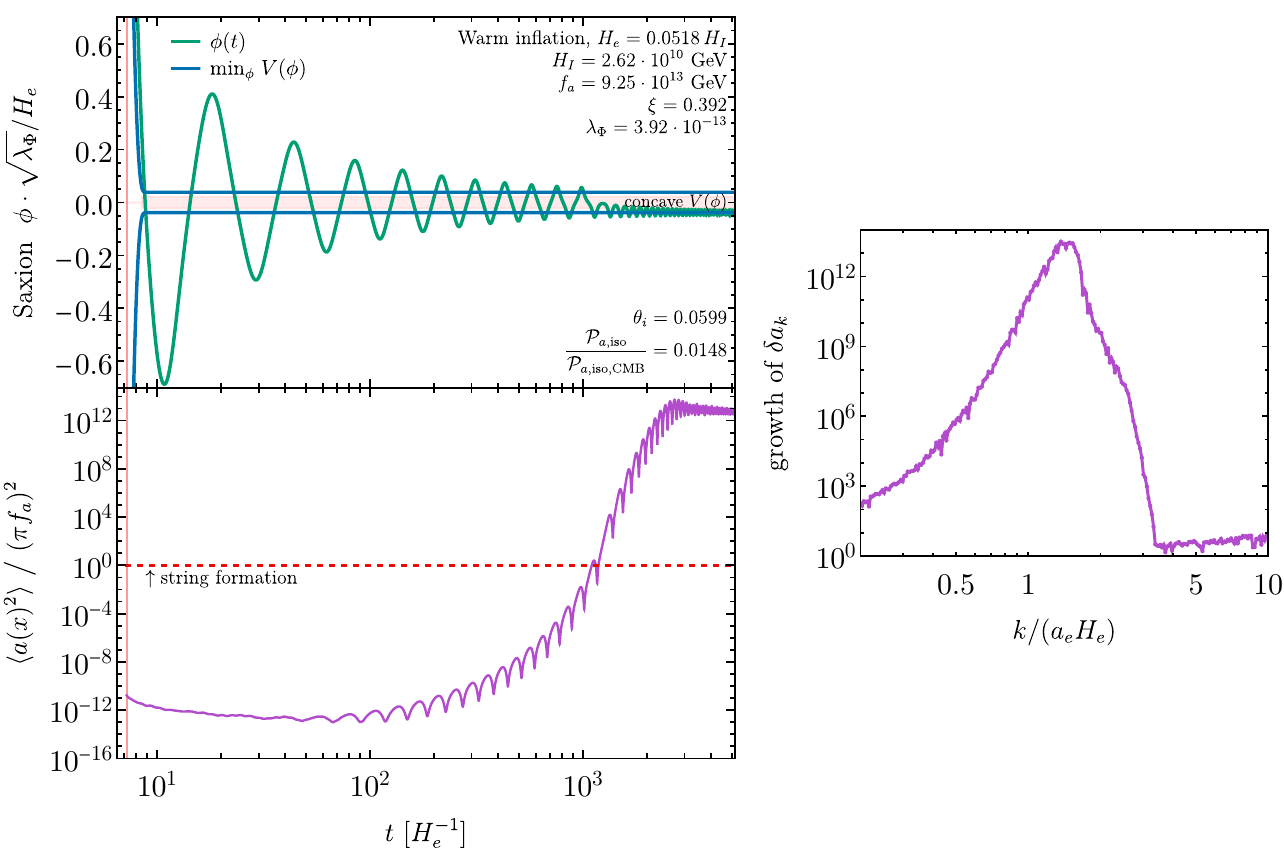}
\caption{Evolution of a point in parameter space displaying a strong parametric resonance. As the \textcolor{ForestGreen}{\textbf{saxion}} (shown in green) oscillates widely around the origin, the \textcolor{Purple}{\textbf{axion}} (shown in purple) direction grows exponentially, until the saxion settles in a minimum and its amplitude decreases.
\textit{Top left}: time evolution of the saxion (radial direction; strictly speaking, this panel shows $\ovX(t)$) after $\te$ (marked by a vertical red line; time is in units of $\He^{-1}$) in green, with the blue lines showing the position of the minima of the potential.
\textit{Right}: Spectrum of axion perturbations due to parametric resonance, computed as the maximum amplitude reached by each mode $y_k \approx \delta a_k (t)$.
\textit{Bottom left}: variance of axion fluctuations (more precisely, we plot the fluctuations $\langle(Y-\ovY)^2\rangle$, which match the axion power spectrum up to a factor $(\phi(t)/\fa)^2$), normalised to $(\pi f_a)^2$. The dashed red line marks the threshold for the formation of axion strings, and the restoration of the PQ symmetry. When the axion passes this threshold, our linearised analysis is no longer a good approximation.
}
\label{fig:warm big par-res}
\end{figure}

\begin{figure}[th]\centering
\includegraphics[width=.9\textwidth]{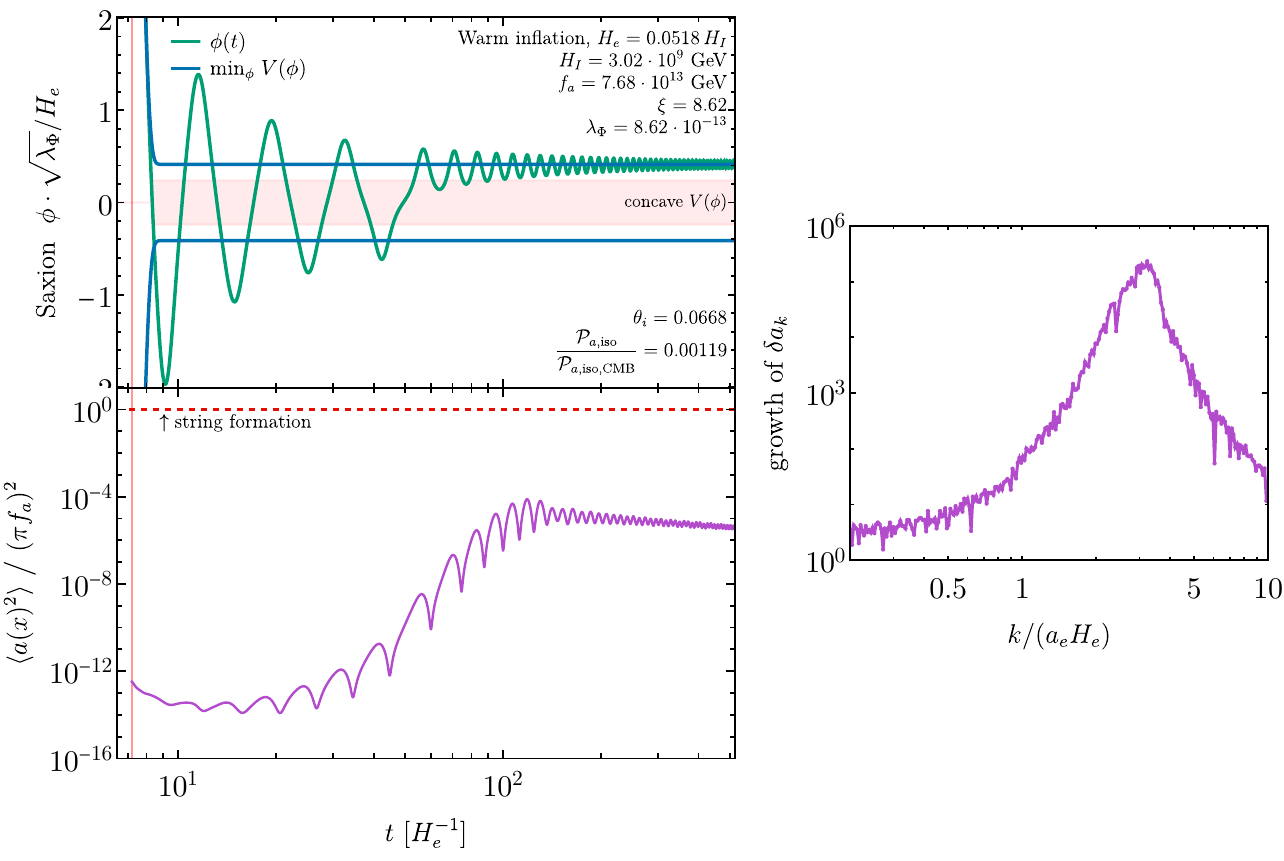}
\caption{Same as Fig.~\ref{fig:warm big par-res}, for a point in parameter space whose evolution does not display a significant parametric resonance. If the saxion swipes no more than $\sim 7-8$ times across the origin before settling in its minimum, the axion does not grow significantly. 
Notice that the parametric resonance stops when the saxion stops reaching the concave part of the wine-bottle potential (marked with a light red band).}
\label{fig:warm small par-res}
\end{figure}

This model-dependent part of the dynamics determines the position and velocity of the saxion background (green line) as visible in the top left corner of \cref{fig:warm big par-res,fig:warm small par-res}.
The blue lines mark the positions of the saxion minima (in the direction of the initial misalignment angle $\thi$) as a function of time. 
For the warm inflation model considered in this section, at the end of inflation they relax exponentially fast to their late-time value $\fa$.

As the wine-bottle potential is restored to its late-time shape at the end of inflation, the saxion is dropped radially from a position that can be well above the tip of the wine-bottle potential. 
This depends both on the initial enhancement $\fainf/\fa$ and on the inflationary model, and sets the number of oscillations of the saxion field before it settles in the valley of its potential.
\cref{fig:warm big par-res} (top left panel) displays 21 such crossings of the origin by the saxion (with $\fainf/\fa\approx 1,390$), whereas \cref{fig:warm small par-res} features  $\fainf/\fa\approx 600$ and 8 such crossings.
In our numerical solution, the saxion field dissipates its kinetic energy through Hubble friction. Other model-dependent sources of friction might be relevant, and assist the settling of the saxion in its minimum, although their contribution to the thermal potential of the saxion might potentially restore the $\UPQ$ symmetry.

In the right-hand side panels of \cref{fig:warm big par-res,fig:warm small par-res}, we plot the maximum enhancement reached by $\sim 250$ modes $y_k$ in the range $(0.2-10)\ke$, with $\ke=\ae \He$, evolved according to the linearised equations of motion \cref{eq:eom Y pert} in the background evolution of $\ovX(t)$. 
We sample three random initial phases for $y_k$ and we average over the corresponding solutions.
The initial amplitude of $y_k$ around Hubble crossing is set by quantum fluctuations, as discussed in \cref{sec:scan}.
As these modes pass through the first instability band of the parametric resonance, they grow exponentially for some time.
The unstable modes, in units of $\ke$, are given by $\phi(t) \cdot \sqrt{\l}/H(t)$, which can be read off by the vertical range of the top left panels.

The power spectrum $\tfrac{1}{\pi^2\fa^2}\int \tfrac{k^3}{2\pi^2} |y_k|^2 \d \ln k=\tfrac{1}{\pi^2}\langle (\theta(x)-\thi)^2\rangle$ is plotted in the bottom left panels of \cref{fig:warm big par-res,fig:warm small par-res}.
As this quantity becomes $\gtrsim \mathcal O(10^{-2})$, our linearised analysis ceases being a good approximation, as the terms that we neglected in deriving \cref{eq:eom XY pert} become large, and the cross-coupling of various modes is non negligible, requiring a lattice study. 
For our scope, a linearised analysis is perfectly suitable to identify a safe range where the non-thermal symmetry restoration is avoided and strings are not formed.
We notice that, for a given $k$, the oscillations of $\delta y_k^2$ do not depend on our choice of the phase of its initial conditions (as we checked explicitly), but they are in phase with the radial mode which is driving the resonance. 
By integrating over various modes in the power spectrum, these synchronous oscillations of different modes average out only partially, as visible in the bottom left panels of \cref{fig:warm big par-res,fig:warm small par-res}, and they are present as long as the parametric resonance is active.

We notice, from the comparison of the top and bottom left panels of \cref{fig:warm big par-res,fig:warm small par-res}, that the parametric resonance occurs for about as long as the saxion field probes the concave part of its potential $V(\phi)$ (shaded in light red in the figures), so that the frequency of some axion fluctuations $y_k$ is negative in \cref{eq:eom Y pert}.

We can try to roughly estimate the growth of the axion by estimating the amount of time that the saxion spends in the concave part of $V(\phi)$ to grasp the amount of growth of the axion modes. 
Each time the saxion crosses the origin, the axion fluctuations (in the appropriate band) undergo an exponential growth, and we checked that, for the points of our scans, one can roughly predict
\begin{equation}
\label{eq:exp growth}
\dthpr \sim \frac{\HI}{2\pi \fa} \exp\left( \mathcal O(1) \cdot n_\text{$\phi$-crossings}\right) \,,
\end{equation}
with $n_\text{$\phi$-crossings}$ defined as the number of times that the saxion crossed the origin (respectively 21 and 8 in \cref{fig:warm big par-res,fig:warm small par-res}).
This number depends in turn on $\fainf/\fa$, as the evolution of $\fainf(t)$ towards the end of inflation determines the initial amplitude of the saxion when it is released above the wine-bottle potential.
This relation is more difficult to estimate analytically though, as it depends on the detailed expression of $\HI(t)$ towards the end of inflation. $n_\text{$\phi$-crossings}$ might also be enhanced from oscillations of $R(t)$ in the preheating phase, before the universe is radiation-dominated and $R\approx0$.

\subsection{Sources of friction for saxion and axion}
\label{app:friction}
In our paper, for simplicity and to be conservative we do not include friction terms for the saxion or the axion due to the couplings with other fields. 
The built-in coupling between saxion and axion is proportional to $\l$ and is thus negligible.
Other couplings might lead to sizeable friction for $\phi$ or $a$, which has a two-fold impact on our story.

As a partial draw-back, particle production might lead to thermal corrections which could potential restore $\UPQ$, thus producing a scenario qualitatively similar to post-inflation with a typical overproduction of axions (for $\fa> 10^{12}\GeV$).

If this does not occur, then the dissipation of the kinetic energy of the saxion at the end of inflation, or a sizeable friction for the axion fluctuations can both assist in suppressing the growth due to parametric resonance, and further enlarge the ``rescued'' parameter space of \cref{fig:isocurvature-rescued-summary}.

In this appendix, we only comment on two possible scenarios of such friction sources. 
A first channel, that is guaranteed for the QCD axion, is the sphaleron friction into SM gluons for the axion \cite{McLerran:1990de,Berghaus:2019whh, DeRocco:2021rzv, Mirbabayi:2022cbt}.
Accounting for the presence of a chiral suppression of the sphaleron emission rate in presence of light quarks \cite{McLerran:1990de,Berghaus:2020ekh,Berghaus:2025dqi} we find that, also due to the decrease of $\as$ at high scales around $T\sim 10^{13}\GeV$, this friction is of order $\Gsph\approx \tfrac{(N_c\as)^5T^3}{2\fa^2}\frac{y_q^2}{2(N_c\as)^4+y_q^2}\approx 10^{-14}T\big(\tfrac{y_q}{10^{-4}}\tfrac{T/\fa}{10^{-2}}\big)^2\ll H$. 
This friction channel is then negligible in a minimal scenario. 

\begin{figure}[h!]\centering
\includegraphics[width=.6\textwidth,trim = 2 570 640 2,clip]{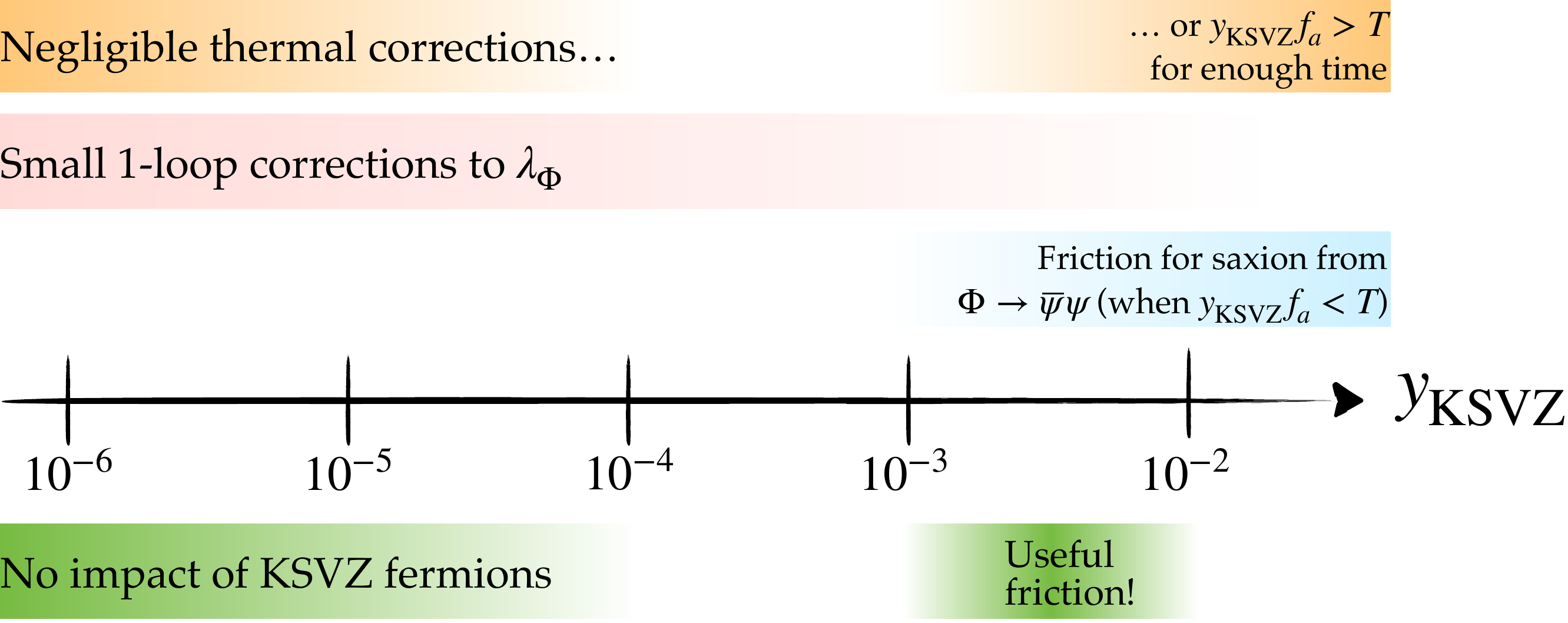}
\caption{Graphical sketch of the interesting range for $\yKSVZ$ where KSVZ fermions might produce a sizable friction for the saxion field without excessive thermal corrections or 1-loop corrections to $\l$.}
\label{fig:Sketch_y-KSVZ}
\end{figure}

Another scenario for which we perform some estimates, summarised in \cref{fig:Sketch_y-KSVZ}, is a KSVZ completion for the axion, with coloured fermions $\psi$ coupled to the saxion field via a Yukawa coupling $\yKSVZ$. We also understand the common assumption that these KSVZ fermions have a hypercharge that allows them to decay into SM quarks, to avoid the risk of overclosure.
The first orange band shows where thermal corrections from a bath of KSVZ fermions do not restore $\UPQ$: their thermal corrections might be subdominant (first range), or too heavy to be produced and stable for enough time to effectively restore the symmetry (second range).
The second pink band quantifies where the 1-loop quantum corrections $\delta\l\sim \tfrac{N_c}{(4\pi)^2}\yKSVZ^4$ are smaller than $\l$, for the points considered in our scan.
The third blue band quantifies the rough range where we found that the decay $\phi\to \overline\psi\psi$ (kinematically open when $\yKSVZ \fa <T$, see the detailed discussion in \cite{Gouttenoire:2021jhk,Eroncel:2024rpe}) could impact appreciably the evolution of the saxion, reducing the growth of the axion modes.
In summary, from our rough estimates there appears to be window of opportunity for $\yKSVZ\sim 10^{-3}-10^{-2}$ where the KSVZ fermions might induce a non-negligible friction for the saxion and reduce the parametric resonance of the axion, while not restoring $\UPQ$.

% !TEX root = Draft_Axion-Isocurvature.tex

\section{Comparison of different inflationary models and couplings to the saxion}
\label{app:infl models}

This Appendix provides further details about the models of inflation that we consider in this paper.
Given the illustrative purpose of our choice of inflationary models, with our focus on the dynamics of the saxion and axion, we do not impose that the model satisfies the constraint on $n_s$ from CMB data. We ensure though that inflation can last for 60 $e$-folds (which is the duration of our numerical evolution), and that $\HI$ is allowed by the bound on $r$.

\begin{figure}[h]\centering
\includegraphics[width=0.5\textwidth,trim = 510 0 0 370, clip]{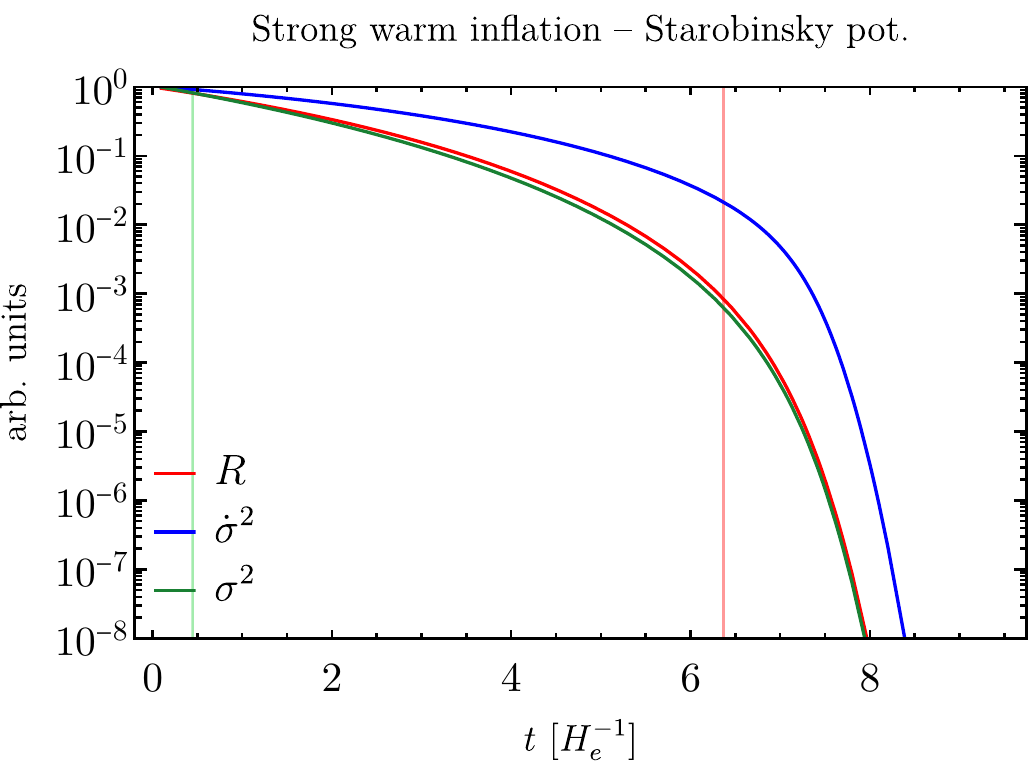}%
\includegraphics[width=0.5\textwidth,trim = 0 0 509 370, clip]{Infl-models_comparison}
\includegraphics[width=\textwidth,trim = 0 370 0 0, clip]{Infl-models_comparison}%
\caption{Evolution of the quantities $(R,\dot \sigma^2,\sigma^2)$ ($\sigma$ being the inflaton field) that can induce a negative mass term for the saxion during inflation (enhancing $\fainf$) through couplings $\xi R|\Phi|^2$, $(\partial_\mu\sigma)^2 |\Phi|^2$, $\sigma^2 |\Phi|^2$.
The quantities are normalised to their initial value. 
The green and red vertical lines in the four plots mark respectively the times at $60$ $e$-folds before $\te$, and $\te$.
The four panels corresponds to four inflationary models: a Warm inflation model, Starobinsky inflation, quadratic and hilltop inflation. 
The models on the top row are the ones for which we performed a scan of the saxion parameter space.}
\label{fig:infl compare}
\end{figure}

The model for which we have displayed most of our results is a \textbf{Warm Inflation} model, as it displays a wider decrease of $\HI(t)$ during the 60 $e$-folds of inflation and the end of inflation displays a smooth transition to radiation domination with $R=0$, without intermediate matter dominated phase where $R$ oscillates (adding a further layer of complexity to the evolution of saxion and axion).
As a concrete implementation, we assume that the inflaton $\sigma$ couples with a strongly coupled Yang-Mills dark sector via a coupling $\sigma G\widetilde G$, inducing a friction for the inflaton due to thermal production of sphalerons \cite{McLerran:1990de, Berghaus:2019whh, DeRocco:2021rzv, Mirbabayi:2022cbt}. 
The effective friction term in the inflaton equations of motion becomes $(3H+\Usph) \dot\sigma$, with 
\begin{equation}
\begin{aligned}
\Usph & = (N_c \alpha)^5 \frac{\TDR^3}{2 f^2}\,, & & N_c =3,\ \alpha = 0.1,\\
\dot \rDR & = -4H\rDR + \Usph \dot \sigma^2\,, & & \rDR= \frac{\pi^2 g_{\star,\textsc{dr}}}{30} \TDR^4 \,, \ g_{\star,\textsc{dr}} = 2(N_c^2-1)\,,
\end{aligned}
\end{equation}
with $f = 1.17 \cdot 10^{13} \GeV\cdot(\HI/10^{10}\GeV)^{1/4}$ chosen to ensure a regime of strong warm inflation with $Q \equiv \Usph /(3\HI) \approx 180$. 
We notice that this value lies within 1-2 orders of magnitude from the axion decay constants $\fa$ of most points of our scan in \cref{fig:warm scan}, which can ease the formulation of a complete model for saxion and inflaton. 
As for the inflationary potential, we use the Starobinsky potential in \cref{eq:starobinsky potential} with $c=0.3$, which was studied in a warm inflation setting in \cite{DeRocco:2021rzv}.
Finally, we notice that the bath at a temperature $\TDR$ might not include the SM species if these are not appreciably coupled to the dark Yang-Mills sector. 
The reheat temperature of the SM and PQ sectors determine the boundary of the post-inflationary region in \cref{fig:warm scan,fig:warm xi-over-lambda}, and it might be significantly smaller than $\TDR(\te)$. 
As a conservative assumption, we take them to coincide, but it is worth remembering that the lower boundary of the ``rescued'' region in \cref{fig:warm xi-over-lambda} might be lowered appreciably.

\begin{figure}[th]\centering
\begin{tikzpicture}%
\hspace{-.07\textwidth}%
\node [above right,inner sep=0] (image) at (0,0) {%
\includegraphics[width=.57\textwidth, trim= 0 0 23 0,clip]{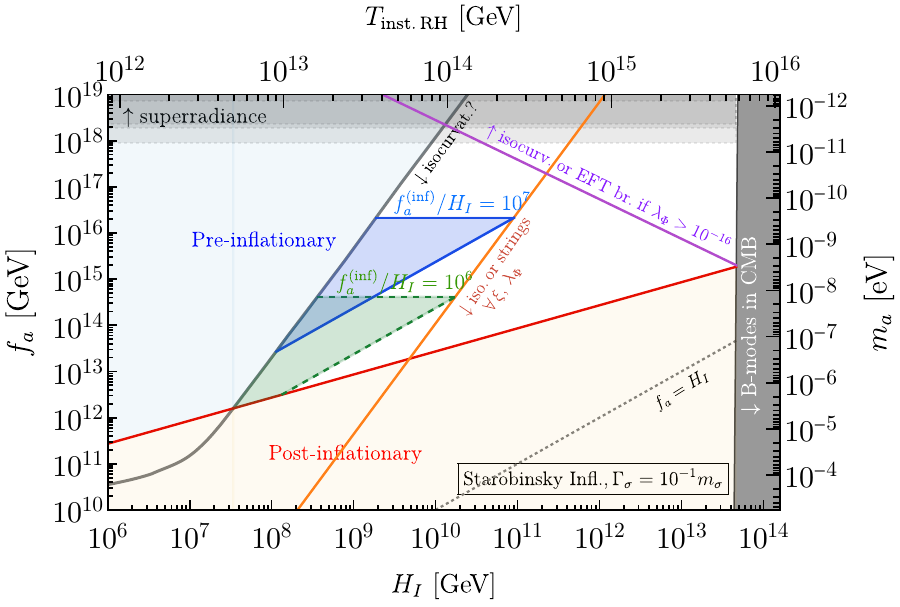}%
\includegraphics[width=.57\textwidth, trim= 23 0 0 0,clip]{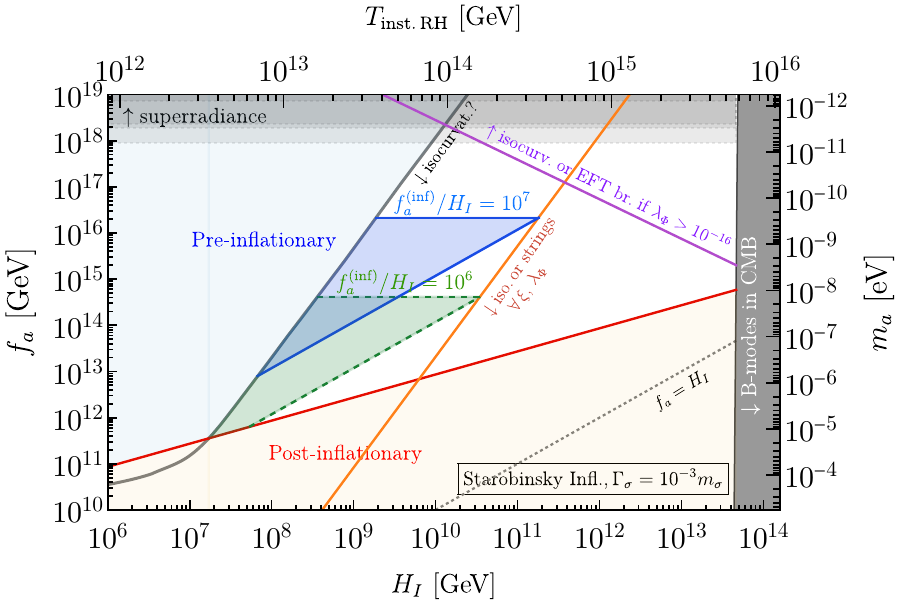}%
};%
\end{tikzpicture}
\vspace{-2em}
\caption{Same as \cref{fig:warm xi-over-lambda}, for the model of Starobinsky inflation with $\Gs=10^{-1}\ms$ (left plot) and $\Gs=10^{-3}\ms$ (right plot).}
\label{fig:cold par space}
\end{figure}

The second model for which we perform a scan of the parameter space is \textbf{Starobinsky inflation} \cite{Starobinsky:1980te} ($R+R^2$), which is also a special case $\alpha=1$ of the $\alpha$-attractor models \cite{Ferrara:2013rsa,Kallosh:2013tua,Kehagias:2013mya}.
We use the scalar-field version of the Starobinsky model (related via a conformal transformation \cite{Whitt:1984pd}), and the inflaton potential reads
\begin{equation}
\label{eq:starobinsky potential}
V_\text{Starobinsky}(\sigma) = 3\HI^2 \MP^2 
\left(1- \exp\left(-c\frac{\sigma}{\MP}\right)\right)^2\,, \quad c= \sqrt{2/3} \,.
\end{equation}
We model the reheating phase with a decay rate $\Gs=10^{-(1,3,5)}\ms$, with $\ms=2\HI$ the inflaton mass in the origin. Starting from the time $\te$, we turn this decay rate on in the equations of motion for the inflaton, with a friction $(3H+\Gs)\dot\sigma$.
By comparing the evolution of $R$ in \cref{fig:infl compare} to the Warm Inflation case, this Starobinsky model displays a shorter and faster excursion of $\HI(t)$ in the last few $e$-folds, which contributes to making the evolution of the saxion less adiabatic and the axion field more subject to parametric resonance (see \cref{fig:isocurvature-rescued}). 
This point, together with the persistence of an oscillating $R(t)$ in the preheating phase, reduce the ratio $R_{\fa}=\fainf/\fa$ to avoid string formation (Eq.~\ref{eq:rescued}) from the value of 700 for Warm Inflation to roughly $(60, 120, 180)$ for $\Gs=10^{-(1,3,5)}\ms$.
The analogue plots of \cref{fig:warm xi-over-lambda} for the Starobinsky model are reproduced in \cref{fig:cold par space}.

In order to exemplify some features of different inflationary models with regard to the evolution of $R(t)$, we also consider in \cref{fig:infl compare} two more models of inflation. 
One is \textbf{quadratic inflation} (or chaotic inflation), the simplest model with just a quadratic term in the inflaton potential.
The inflaton mass to sustain $N_\textsc{cmb}$ $e$-folds of inflation is $\ms = \sqrt{3/(2N_\textsc{cmb})}\HI$.
The last model we plot in  \cref{fig:infl compare} is \textbf{hilltop inflation}, which features a very flat potential around the origin and a fairly large mass around the minimum, with $V(\sigma) =\tfrac{\ms^2 v^2}{72}(1-\sigma^2/v^2)^6$ and $v=\tfrac{\MP}{2}, \, \ms=12\sqrt{6} \HI$.

The comparison of the evolution of $R,\,\sigma^2,\,\dot\sigma^2$ in these models shows that the three variables typically display a qualitatively similar evolution. 
The choice of any of these couplings to $|\Phi|^2$ should then lead to similar conclusions, with upper limits on the size of these couplings set by similar physical arguments (validity of an EFT expansion, unitarity constraints, or flatness or the inflaton potential for the $\sigma^2$ coupling).
The choice of the inflationary model plays instead a bigger role, as models with a smoother and more adiabatic evolution of $\HI(t)$ should help in suppressing the parametric resonance growth for the axion.
The model of warm inflation displays the most gradual transition among them, and additionally features an (exponentially) fast drop of $R(t),\sigma(t)$ at reheating, without an oscillating phase for these terms that occasionally contributes to driving the resonance for the axion further.

%%%%%%%%%%%%%%%%%%%%%%%%%%%%%%%%%%%%%%%%%%

\bibliographystyle{JHEP}
\bibliography{bib_axion-iso.bib}

\end{document}